\documentclass[final,5p,times,twocolumn]{elsarticle}
\UseRawInputEncoding

\usepackage{amssymb}
\usepackage{xcolor}
\usepackage{multirow}
\usepackage{float}
\usepackage{stfloats} 
\usepackage{accents}

\usepackage{hyperref}
\hypersetup{%
    colorlinks=true,        
    linkcolor=blue,          
    citecolor=blue,         
    urlcolor=blue           
    }

\pdfoutput=1

\journal{Physics Letters B}

\begin{document}

\begin{frontmatter}



\title{Constraints for  the X17 boson from compacts objects observations}


\author[lebel1]{A. Kanakis-Pegios}
\ead{alkanaki@auth.gr}
\author[lebel2]{V. Petousis}
\ead{vlasios.petousis@cvut.cz}
\author[lebel2]{M. Veselsk\'y}
\ead{Martin.Veselsky@cvut.cz}
\author[lebel3]{Jozef Leja}
\ead{jozef.leja@stuba.sk}
\author[lebel1]{Ch. C. Moustakidis}
\ead{moustaki@auth.gr}


\affiliation[lebel1]{organization={Department of Theoretical Physics, Aristotle University of Thessaloniki},
            city={Thessaloniki},
            postcode={54124}, 
            country={Greece}}
\affiliation[lebel2]{organization={Institute of Experimental and Applied Physics, Czech Technical University},
            city={Prague},
            postcode={110 00}, 
            country={Czechia}}

\affiliation[lebel3]{organization={Faculty of Mechanical Engineering - Slovak University of Technology in Bratislava},
            city={Bratislava},
            postcode={812 31}, 
            country={Slovakia}}

\begin{abstract}
We investigate the hypothetical X17 boson on neutron stars and Quark Stars (QSs) using various hadronic Equation of States (EoSs) with phenomenological or microscopic origin. Our aim is to set realistic constraints on its coupling constant and the mass scaling, with respect to causality and various possible upper mass limits and the dimensionless tidal deformability $\Lambda_{1.4}$.
In particular, we pay special attention on two main phenomenological parameters of the X17, the one is related to the coupling constant $\mathrm{g}$ that it has with hadrons or quarks and the other with the in-medium effects through the regulator $\mathrm{C}$. Both are very crucial concerning the contribution on the total energy density and pressure.
In the case of considering the X17 as a carrier of nuclear force in Relativistic Mean Field (RMF) theory, an admixture into vector boson segment was constrained by 20\% and 30\%. 
In our investigation, we came to the general conclusion that the effect of the hypothetical X17 both on neutron and QSs constrained mainly by the causality limit, which is a specific property of each EoS. Moreover, it  depends  on the interplay between  the main two parameters that is the interaction coupling $\mathrm{g}$ and the in-medium effects regulator $\mathrm{C}$. These effects are more pronounced in the case of QSs concerning all the bulk properties.

\end{abstract}

\end{frontmatter}



\section{Introduction}
In 2016 an article of Krasznahorkay et al. appeared \cite{Attila1}, where an anomaly in angular correlation of $e^{-}e^{+}$ decay of the 1$^{+}$ excited level of $^{8}$Be nucleus at ${18.15~}$MeV reported, and specifically observed enhancement at folding angles. Since first report till today, Krasznahorkay and his group, reported in addition the same anomaly - observed in the angular correlation of the $e^{-}e^{+}$ emission - in the excited states of ${^4}$He and $^{12}$C \cite{Attila2, Attila4He, Attila22}. The reported anomalies at a folding angles was interpreted as a signature of a new neutral boson with a mass of about ${m_{X}=17~}$MeV.

These reported observations placed the hypothetical X17 boson as a dark matter candidate and in that spirit since then, several theoretical works pursued this claim \cite{Feng1, Feng2}. However, an explanation relating this particle to the QCD vacuum was also proposed \cite{VPL}, in the conjecture that the 17~MeV particle could mediate the nucleon-nucleon interactions at large distances in an unbound cluster configuration. Since the assumption that the 17~MeV boson is the only carrier of nuclear interactions is somewhat extreme, we investigated the possible influence of the hypothetical 17~MeV boson on nuclear matter and its influence on the structure of the compact astrophysical objects like neutron stars \cite{VPL-NS}.

A further investigation \cite{Symmetry}, explores the hypothetical 17~MeV boson in the frame of the Relativistic Mean Field (RMF) theory, constructing a universal Equation of State (EoS) that satisfy all of the well known experimental constraints, from finite nuclei, heavy ion collisions all the way to the neutron stars, allowing a reproduction in masses from ${\cong 1.4\;\mathrm{M_\odot}}$ up to ${\cong 2.5\;\mathrm{M_\odot}}$. The values of the radius show an agreement with the recent measurement by NICER \cite{NICER8,NICER9}. Also the value of the maximum mass is in a good agreement with the recently reported mass of pulsar $2.35\;\mathrm{M_\odot}$ \cite{2.35SM} and potentially also with the mass of the secondary component of the gravitational wave event GW190814 \cite{GW190814}. A previous investigation \cite{Sulaksono}, following an alternative direction, tried to set the ratio ${g{^2}}$/${\mu{^2}}$ for a Weakly Interactive Light Boson (WILB) inside the neutron star. In this investigation, the range of the ratio ${g{^2}}$/${\mu{^2}}$ was estimated to be less than 2~GeV$^{-2}$. A value that looks to be suitable for the neutron star environment talking in account also the experimental constraints for consistency of symmetry and binding energy. Also using the fact that the presence of WILB does not effect the crust properties of neutron star matter they obtaining a quite restricted constraint of the WILB characteristic scale. They also show that the in-medium modification effect indeed cannot be neglected and for their investigation they assumed that the WILB mass follows the same scaling as the one of Brown–Rho\cite{BR}.

Another similar research \cite{Krivoruchenko-2009}, provides a ratio ${g{^2}}$/${\mu{^2}}$ of a WILB to be less than 50~GeV$^{-2}$ and above than 25~GeV$^{-2}$. They also investigated an upper value of ${g{^2}}$/${\mu{^2}}$ = 100~GeV$^{-2}$ which deviates a lot from the must fulfilled restrictions concerning the binding energy and the symmetry energy, even though reproducing an acceptable mass according to the GW190814 observation \cite{GW190814} close to ${\cong 2.5\;\mathrm{M_\odot}}$, which is comparable with the astrophysical observation. In the case of Quark Star (QS) a research \cite{QS_Weber}, considering the non-Newtonian gravity effects, for current quark masses of $m_{u}$ = 2.16~MeV, $m_{d}$ = 4.67~MeV, and $m_{s}$ = 93~MeV, they estimate a range of ${g{^2}}$/${\mu{^2}}$ between 4.58~GeV$^{-2}$ and 9.32~GeV$^{-2}$ reaching a maximum mass ${\cong 2.4\;\mathrm{M_\odot}}$.

In all the aforementioned investigations, the ratio ${g{^2}}$/${\mu{^2}}$ for a WILB and only the ratio, was thoroughly investigated. No attempt was made to estimate separately the coupling $\mathrm{g}$ from the boson mass $\mu$ for the WILB, and that was reasonable, because the mass of a possible candidate boson was not known or guessed at that time. 

In general, concerning the WILB case scenario inside a neutron star and it's connection with the hypothetical U boson, which in our case could be also represented by the hypothetical X17 boson, has been investigated also in the past~\cite{Fujii-1988, Yu-2018, Zhang-2016, Pi-2022, Dong-Zhang-2011,YanXU-2023, Yang-2023, Berryman-2021}.

In our present work, having as a candidate a 17~MeV boson (hereafter X17), we investigating the effects of the non-Newtonian gravity together with a series of different models - hadronic EoSs with phenomenological and realistic or microscopic origin, from the RMF to the Momentum Dependent Interaction (MDI) model and QSs. Our main motivation is to set realistic constraints on the X17 coupling constant $\mathrm{g}$ and the scaling of its mass, which is affected by the changes in the baryon density well above its saturation value inside the neutron star and QS. All of our research is compared with astrophysical observations, reported from LIGO-VIRGO and pulsar observations.

The paper is organized as follows: in Section 2, we briefly describe the non-Newtonian gravity model while  in Section 3, we present the nuclear models in the context of which we are investigating our constraints for the X17 boson. In Section 4, we present the basic formalism of the  tidal deformability  while  in Section 5, we display  and discuss the  results of the present study. In  Section 6, we finalize with our concluding remarks.

\section{The non-Newtonian gravity model}
The deviation of the Newton's gravitational potential, known as non-Newtonian gravity \cite{non-Newtonian}, usually parameterized in the form: 
\begin{equation}
 V(r)=-\frac{Gm_1m_2}{r}\left(1+\alpha_{\tiny {G}} e^{-r/\lambda} \right) = V_{N}(r) + V_{Y}(r)   
 \label{Newt-low)}
\end{equation}
where $V_{N}(r)$ is the Newtonian potential, $V_{Y}(r)$ is the Yukawa correction, $G = 6.67\times10^{-11}$ N m$^{2}$/kg$^{2}$, is the universal gravitational constant, $\alpha_{\rm G}$ is the dimensionless
coupling constant of the Yukawa force and $\lambda$ represents the range of the Yukawa force mediated by the exchange of a boson with mass $\mu$. The above quantities related according to the following relations:
\begin{equation} 
\alpha_{\rm G}=\pm\frac{{\rm g}^2 \hbar c}{4\pi G m_b^2}, \qquad \lambda = \frac{\hbar}{\mu c}
\end{equation}
where the $\pm$ sign refers to scalar(+) and vector(-) boson, $\mathrm{g}$ is the boson-baryon coupling constant and $m_{b}$ is the baryon mass. 


\section{The nuclear models}
\subsection{The Relativistic Mean Field (RMF) model}
In the case of the RMF theory using the extended Dirac-Hartree approximation, the energy density and pressure of neutron matter is given by the following expressions \cite{Serot-1997}:

\begin{eqnarray}
 {\cal E}&=&\frac{(\hbar c)^3g_v^2}{2(m_vc^2)^2}n_b^2+ \frac{(\hbar c)^3  (\frac{g_\rho}{2}){^2}}  {2(m_\rho c^2)^2}\rho_I^2+ \frac{(m_sc^2)^2}{2g_s^2(\hbar c)^3}(m_bc^2-m_b^*c^2)^2\nonumber\\
 &+& \frac{\kappa}{6g_s^3}(m_bc^2-m_b^*c^2)^3+\frac{\lambda}{24g_s^4}(m_bc^2-m_b^*c^2)^4\nonumber\\
 &+&\sum_{i=n,p} \frac{\gamma}{(2\pi)^3}\int_0^{k_Fi} 4\pi k^2 \sqrt{(\hbar c k)^2+(m_i^* c^2)^2}dk
 \label{RMF-E-1}
\end{eqnarray}
\begin{eqnarray}
P&=&\frac{(\hbar c)^3g_v^2}{2(m_vc^2)^2}n_b^2+ \frac{(\hbar c)^3  (\frac{g_\rho}{2}){^2}} {2(m_\rho c^2)^2}\rho_I^2 - \frac{(m_sc^2)^2}{2g_s^2(\hbar c)^3}(m_bc^2-m_b^*c^2)^2\nonumber\\
 &+& \frac{\kappa}{6g_s^3}(m_bc^2-m_b^*c^2)^3+\frac{\lambda}{24g_s^4}(m_bc^2-m_b^*c^2)^4\nonumber\\
 &+&\sum_{i=n,p}  \frac{1}{3}\frac{\gamma}{(2\pi)^3}\int_0^{k_Fi} \frac{4\pi k^2}{\sqrt{(\hbar c k)^2+(m_i^* c^2)^2}}dk
 \label{RMF-E-1}
\end{eqnarray}
where ${\cal E}$ is the energy density, $P$ is the pressure, $g_s$, $g_v$ and $g_\rho$ are the couplings of the scalar boson, vector boson, and iso-vector $\rho$-meson respectively, $m_s$, $m_v$ and $m_\rho$ are the rest masses of scalar and vector bosons and $\rho$-meson respectively, the term $\rho_I$ involves the difference between the proton and neutron densities (important for finite nuclei), also $\kappa$ and $\lambda$ are the couplings of the cubic and quartic self-interaction of the scalar boson, $m_b$ and $m_b^*$ are the rest mass and the effective mass of the nucleon,
$n_b$ is the nucleonic density, $k_F$ is the Fermi momentum of nucleons at zero temperature and $\gamma$ is the degeneracy, with value $\gamma= 4$ for symmetric nuclear matter and $\gamma= 2$ for neutron matter (used in this investigation). 

Considering the possibility that the hypothetical 17~MeV boson \cite{Attila1}, could contribute as a second vector boson, we can write an effective mass term in the following form \cite{Symmetry}: 

\begin{equation}
 m_v^{*2}=a_X^2m_{X}^2+(1-a_X)^2m_{\omega}^2 \label{mass-mix}  
\end{equation}
where $a_X$ is the admixture coefficient of the $m_X = 17$ MeV boson to the total vector potential. Depending on the value of $a_X$, the effective mass can range from $m_{\omega} = 782.5$ MeV to 17 MeV.

Using two different values for the admixture coefficient ($a_X$): 0.2(20\%, $m^*_v$ = 626 MeV) and 0.3(30\%, $m^*_v$ = 547.8 MeV) and increasing the value on the standard $\rho$-coupling by 5\% and 10\% (effective $\rho$-coupling: g$^*_{\rho}$), respectively, we constructed a set of 13 EoSs, fulfilling experimental constraints in analogy to \cite{Symmetry}, where properties of nuclear matter and finite nuclei were considered. 

All combinations and parameters that were used are shown in Table.\ref{tab1} and Table.\ref{tab2}. The corresponding Mass-Radius diagrams are depicted in Fig.\ref{fig:mr_1}.

\begin{center}
\begin{table}[htp]
\caption{Constrained parameter sets for eight EoSs with incompressibilities K$_0$=245-260 MeV, $a_X$ = 0.2 (20\%, $m^*_v$ = 626 MeV), standard $\rho$-coupling = 4.47~\cite{Ring-1997} and effective $\rho$-coupling g$^*_{\rho}$ increased (+) by 5\% and 10\% compared to its standard value.}
\vskip 0.2cm
{\centering \small
\begin{tabular}{lllllllr}
EoS&g$^*_{\rho}$&g$_{v}$&g$_{s}$&m$_{s}[MeV]$&$\kappa$[MeV]&$\lambda$ \\
\hline\hline
\vspace{-0.3cm}
& & & & & & & \\
E1&+5\%&7.61&6.78&406.6&19.0&-60.0 \\
E2&+5\%&8.00&6.76&391.4&17.0&-63.3 \\
E3&+5\%&8.00&7.03&405.6&19.5&-80.0 \\
\hline
E4&+10\%&7.23&7.27&451.9&25.0&-33.3 \\
E5&+10\%&7.23&7.27&451.9&25.5&-46.7 \\
E6&+10\%&7.23&7.51&467.0&28.5&-56.7 \\
E7&+10\%&7.61&7.03&421.7&21.0&-60.0 \\
E8&+10\%&7.61&7.03&421.7&21.5&-73.3 \\
\hline\hline
\end{tabular}
\par}
\label{tab1}
\end{table}
\end{center}

\begin{center}
\begin{table}[htp]
\caption{Constrained parameter sets for five EoSs with incompressibilities K$_0$=245-260 MeV, $a_X$ = 0.3 (30\%, $m^*_v$ = 547.8 MeV), standard $\rho$-coupling = 4.47~\cite{Ring-1997} and effective $\rho$-coupling g$^*_{\rho}$ increased (+) by 5\% and 10\% compared to its standard value.}
\vskip 0.2cm
{\centering \small
\begin{tabular}{lllllllr}
EoS&g$^*_{\rho}$&g$_{v}$&g$_{s}$&m$_{s}$[MeV]&$\kappa$[MeV]&$\lambda$ \\
\hline\hline
\vspace{-0.3cm}
& & & & & & & \\
E9&+5\%&7.61&8.08&451.9&19.0&-103.3 \\
E10&+5\%&8.00&8.35&451.9&18.5&-123.3 \\
E11&+5\%&7.23&8.33&482.2&26.0&-150.0 \\
\hline
E12&+10\%&6.47&5.77&346.1&14.5&-33.3 \\
E13&+10\%&7.23&8.07&467.0&23.0&-123.3 \\
\hline\hline
\end{tabular}
\par}
\label{tab2}
\end{table}
\end{center}

\subsection{The Momentum Dependent Interaction (MDI) model}
The Momentum Dependent Interaction (MDI) model used
here, was already presented and analyzed in a previous paper~\cite{Prakash-1997,Moustakidis-2009}. The MDI is designed to reproduce the results of the microscopic calculations of both nuclear and neutron rich matter at zero temperature and can be extended to finite
temperature. The energy density of the baryonic matter, is given by:
\begin{eqnarray}
{\cal E}(u,I)&=&\frac{3}{10}E_{F}^{0}n_0\left[(1+I)^{5/3}+(1-I)^{5/3}\right]u^{5/3}\nonumber \\
&+&
\frac{1}{3}{\cal A}n_0\left[\frac{3}{2} -\left(\frac{1}{2}+x_0   \right)I^2 \right]u^2  \nonumber \\
&+&\frac{\frac{2}{3} {\cal B} n_0\left[\frac{3}{2} -\left(\frac{1}{2}+x_3   \right)I^2 \right]u^{\sigma+1} }{1+\frac{2}{3} {\cal B}' n_0\left[\frac{3}{2} -\left(\frac{1}{2}+x_3   \right)I^2 \right]u^{\sigma-1}} \nonumber \\
&+& u\sum_{i=1,2}\left[\frac{}{} {\cal C}_i ({\cal J}_n^i+{\cal J}_p^i)+
\frac{( {\cal C}_i-8Z_i)}{5}I({\cal J}_n^i-{\cal J}_p^i)\right] 
\label{ED-MDI}
\end{eqnarray}
In Eq.~(\ref{ED-MDI}), $E_{F}^{0}$ is the Fermi energy of symmetric nuclear matter
at the equilibrium density $n_0 =
0.16$ fm$^{-3}$,  $I = (n_n-n_p)/n$ and $u = n/n_0$. The parameters ${\cal A}$, ${\cal B}$, $\sigma$, ${\cal C}_1$, ${\cal C}_2$, and ${\cal B}'$ which
appear in the description of symmetric nuclear matter and the
additional parameters $x_0$, $x_3$, $Z_1$, and $Z_2$ used to determine
the properties of asymmetric nuclear matter, are treated as
parameters constrained by empirical knowledge~\cite{Prakash-1997}. By suitable choice of the above parameters we can regulate the stiffness of the corresponding EoS. This stiffness is well reflected by the values of the slope parameter $L$ which defined as: 
\begin{equation}
 L=3n_0\left(\frac{\partial E_{sym}(n)}{\partial n}  \right)_{n=n_0}   
 \label{slope-1}
\end{equation}
where the symmetry energy, in general is defined as, 
\begin{equation}
 E_{sym}(n)=\frac{1}{2!}\left(\frac{\partial^2E(n,I)}{\partial I^2}  \right)_{I=0}   
\end{equation}
and $E(n,I)={\cal E}(u,I)/n$ is the energy per baryon. 
Moreover, the quantity ${\cal J}_{\tau}^i (n,I,T)$ is defined as: 
\begin{equation}
{\cal J}_{\tau}^i (n,I,T)=2\int \frac{d^3k}{(2\pi)^3}{\rm g}(n,\Lambda_i)f_{\tau}   
\end{equation}
where $f_{\tau} $, (for $\tau$ = n, p) is the Fermi-Dirac distribution function. The function $ {\rm {\cal G}}(k,\Lambda_i)$ ) suitably chosen to simulate finite range
effects is of the following form:
\begin{equation}
 {\rm {\cal G}}(k,\Lambda_i)=\left[1+\left(\frac{k}{\Lambda_i}  \right)^2 \right]^{-1}   
\end{equation}
where the finite-range parameters
are $\Lambda_1=1.5 k_F^0$  and $\Lambda_2=3 k_F^0$   and $k_F^0$ is the Fermi
momentum at the saturation point $n_0$.

\subsection{Contribution on energy and pressure of the X17 boson}
The energy density of the WILB boson in neutron star matter is given by~\cite{Krivoruchenko-2009,Wen-2009}:
\begin{equation}
{\cal E}_{\rm B}=\pm\frac{(\hbar c)^3}{2}\left(\frac{{\rm g}}{m_Bc^2}  \right)^2n_b^2
\label{En-pre-1}
\end{equation}
Where ${\rm g}$  is the coupling constant of the interaction and $m_Bc^2 $ the mass of the boson. The sign (+) corresponds to a vector boson (repulsive interaction) and (-) to a scalar boson (attractive interaction). 
The energy per baryon then is given by $E_{bar}={\cal E}_{\rm B}/n_b$.

The corresponding contribution  on the pressure is defined as:
\begin{eqnarray}
P_B &=& n^2\frac{\partial ({\cal E_B}/n)}{dn} \nonumber \\
&=& \frac{(\hbar c)^3}{2}\left(\frac{{\rm g}}{m_Bc^2}  \right)^2n_b^2\left(1-\frac{2n_b}{m_Bc^2}\frac{\partial (m_Bc^2)}{\partial n_b} \right)
\label{Pres-neutron}
\end{eqnarray}
In the specific case where the the mass does not depends on the density the  pressure contribution is identical to that of the energy density.
Now, the total equation of state is just the sum of the baryons and bosons that is:
\begin{equation}
{\cal E}={\cal E}_{\rm bar}\pm {\cal E}_{\rm B}, \quad P=P_{\rm bar}\pm P_{\rm B} 
\end{equation}
In previous work \cite{Sulaksono,Krivoruchenko-2009,Wen-2009} the range of the ratio $\left({\rm g}/{m_Bc^2}\right)^2$ was varying between $[0-200]\;{\rm GeV}^{-2}$. However, in the present study we consider that the coupling varies in the interval $[0.01-0.022]$ which corresponds (for $m_Bc^2=17$ MeV) the interval for $(g/m_Bc^2)^2$ $[0.346,1.675]$ GeV$^{-2}$.

According to Brown and Rho~\cite{BR}, the in-medium modification of the mesons follows the linear scaling:
\begin{equation}
m_{\rm B}^*\equiv m_{\rm B}\left(1-C\frac{n_b}{n_0} \right)
\ ({\rm MeV})
\label{mass-medium}
\end{equation}

We consider, following the suggestion in Ref.\cite{Sulaksono} that, at least at low densities, in-medium modification of X17 mass takes  a similar form to Eq.\ref{mass-medium}. The parameter $\mathrm{C}$ is fixed in order the predicted EoS to be compatible with the bulk properties in symmetric nuclear matter ($E_{\rm bind}=-16$ MeV and $n_0=0.16$ fm$^{-3}$).

A suitable parametrization is also the following:
\begin{equation}
 m_{\rm B}^*\equiv \frac{17}{1-C}\left(1-C\frac{n_b}{n_0}  \right) \ ({\rm MeV})
\label{mass-medium-2}   
\end{equation}
which ensures that at the saturation density $n_0$, $m_{\rm B}^*=17$ MeV. 
To good approximation the contribution of the WILB begins after the crust-core interface. 

In previous studies, it was found that in the case of vector boson, the total equation of state becomes very stiff leading to a very large values of the maximum mass and corresponding larges radius. On the other hand in the case of scalar boson the EoS is so soft leading to a very small values (out of the observations) of the maximum mass. In any case the majority of the studies focuses only on the maximum mass and not on the radius and the tidal deformability. The aforementioned quantity directly related with information obtained from the detection of gravitational waves.  

\subsection{Quark Stars (QSs)}
In the present work we use  the Color-Flavor Locked (CFL) model for the Quark Stars (QSs), following the previous work of Lugones and Horvath~\cite{Lugones-2002}. In particular, we use the lowest order approximation, in which the EoSs are given in a simple and analytical form and in particular the contribution of the pairing correlations and the mass of the strange quark are provided explicitly. The pressure, energy density and the baryon density are given by the expressions~\cite{Lugones-2002} (see also~\cite{Roupas-2021}):
\begin{equation}
P_Q=\frac{3\mu^4}{4\pi^2(\hbar c)^3}-\frac{3(m_sc^2)^2\mu^2}{4\pi^2(\hbar c)^3}+\frac{3\Delta^2\mu^2}{\pi^2(\hbar c)^3}-B
\label{Pres-quark}
\end{equation}
\begin{equation}
{\cal E}_Q=\frac{9\mu^4}{4\pi^2(\hbar c)^3}-\frac{3(m_sc^2)^2\mu^2}{4\pi^2(\hbar c)^3}+\frac{3\Delta^2\mu^2}{\pi^2(\hbar c)^3}+B
\label{Energ-quark}
\end{equation}
and 
\begin{equation}
n_b=\frac{\mu^3}{\pi^2(\hbar c)^3}-\frac{(m_sc^2)^2\mu}{2\pi^2(\hbar c)^3}+\frac{2\Delta^2\mu}{\pi^2(\hbar c)^3}=\frac{\mu^3}{\pi^2(\hbar c)^3}+\frac{3\mu \alpha}{\pi^2 (\hbar c)^3}
\label{dens-quark}
\end{equation}
where 
\begin{equation}
\mu^2=-3\alpha+\sqrt{9\alpha^2+\frac{4}{3}\pi^2(P_Q+B)(\hbar c)^3  }  
\label{mu-2}
\end{equation}

and
\begin{equation}
\alpha=-\frac{(m_sc^2)^2}{6}+\frac{2\Delta^2}{3}  
\label{alpha}
\end{equation}
In this study we use the parametrization of the EoS according to the  paper~\cite{Flores-2017}. In particular each equation of state is denoted as CFLX[$B$,$\Delta$,$m_s$], where $X$, is the numbering of the specific EoS, $B$ (in MeV fm$^{-3}$), $\Delta$ (in MeV ), $m_s$ (in MeV) are the bag constant, the pairing and the mass of the strange quark respectively. It is worthwhile to notice here that, in general,  the lower the value of $B$ and $m_s$ the stiffer the EoS as well as the lower values of $\Delta$ the softer the EoS.  In the present study we use the sets
CFL2[60,50,150] (intermediate stiffness), CFL10[80,150,150] (stiff) and CFL13[100,100,150] (soft) in order to cover a large region of stiffness.   

The energy density of the bosons in the case of quark stars is given by~\cite{QS_Weber,Yang-2023}: 
\begin{equation}
{\cal E}_{\rm B} = \pm\frac{9(\hbar c)^3}{2}\left(\frac{{\rm g}}{m_Bc^2}  \right)^2n_b^2
\label{En-pre-1-quark}
\end{equation}
where the contribution of the pressure is given by an expression similar to (\ref{Pres-neutron}). 
The total equation of state is just the sum of the quarks and bosons that is:
\begin{equation}
 {\cal E}={\cal E}_{\rm Q}\pm {\cal E}_{\rm B}, \quad P=P_{\rm Q}\pm P_{\rm B} 
\end{equation}
It is worth clarifying here that the main difference with the case of neutron stars is the extra factor of 9 in terms of the contribution to pressure and energy density. This additional contribution plays, as we shall see, an important role in the effect of the boson on the basic properties of compact objects.

\subsection{Speed of sound}
The speed of sound is a very crucial quantity since is directly related with the stiffness of the EoS. In particular, the adiabatic speed of sound is defined as: 
\begin{equation}
v_s/c=\sqrt{\left(\frac{d P}{d {\cal E}}\right)_S}  
\label{speed-1}
\end{equation}
where $S$ is the entropy per baryon. More importantly, the speed of sound  introduce an upper limit on stiffness of the EoS according to which the speed of sound cannot exceed the speed of light. Bedaque and Steiner \cite{Bedaque-2015} have provided simple
arguments that support as an upper limit the value  $c/\sqrt{3}$ in nonrelativistic and/or weakly coupled theories. These authors
pointed out that the existence of neutron stars with masses
about two solar masses combined with the knowledge of the
EoS of hadronic matter at low densities is not consistent with
this bound. In any case, in studies related with the predictions of bulk properties of compact object special attention must be given on the density dependence of the speed of sound and the upper limits must be carefully  taken into account.

\section{Tidal Deformability}
Very important sources of gravitational waves are those produced during the phase of the inspiral on a binary system of neutron stars before they finally merge~\cite{Postnikov-2010,Flanagan-08,Hinderer-08}. These kind of sources leads to the measurement of various properties of neutron stars. During the inspiral phase the tidal effects can be detected~\cite{Flanagan-08}.

The $k_2$ parameter, also known as tidal Love number, depends on the EoS and describes the response of a neutron star to the tidal field $E_{ij}$ with respect to the induced quadrupole field $Q_{ij}$~\cite{Flanagan-08,Hinderer-08}. Their exact relation is given below:
\begin{equation}
Q_{ij}=-\frac{2}{3}k_2\frac{R^5}{G}E_{ij}\equiv- \lambda E_{ij},
\label{Love-1}
\end{equation}
where $R$ is the neutron star radius and $\lambda=2R^5k_2/3G$ is the tidal deformability.
Also, another quantity that is well measured is the effective tidal deformability $\tilde{\Lambda}$ which is given by~\cite{Abbott-2019-X}
\begin{equation}
\tilde{\Lambda}=\frac{16}{13}\frac{(12q+1)\Lambda_1+(12+q)q^4\Lambda_2}{(1+q)^5},
\label{L-tild-1}
\end{equation}
where the mass ratio $q=m_2/m_1$ lies within the range $0\leq q\leq1$
and $\Lambda_i$ is the dimensionless deformability~\cite{Abbott-2019-X}
\begin{equation}
\Lambda_i=\frac{2}{3}k_2\left(\frac{R_i c^2}{M_i G}  \right)^5\equiv\frac{2}{3}k_2 \beta_i^{-5}  , \quad i=1,2.
\label{Lamb-1}
\end{equation}
The effective tidal deformability $\tilde{\Lambda}$ plays important role on the neutron star merger process and is one of the main quantities that can be inferred by the detection of the corresponding gravitation waves. 
\section{Results and Discussion}
\begin{figure}[h]
    \centering
    \includegraphics[width=80mm, height=70mm]{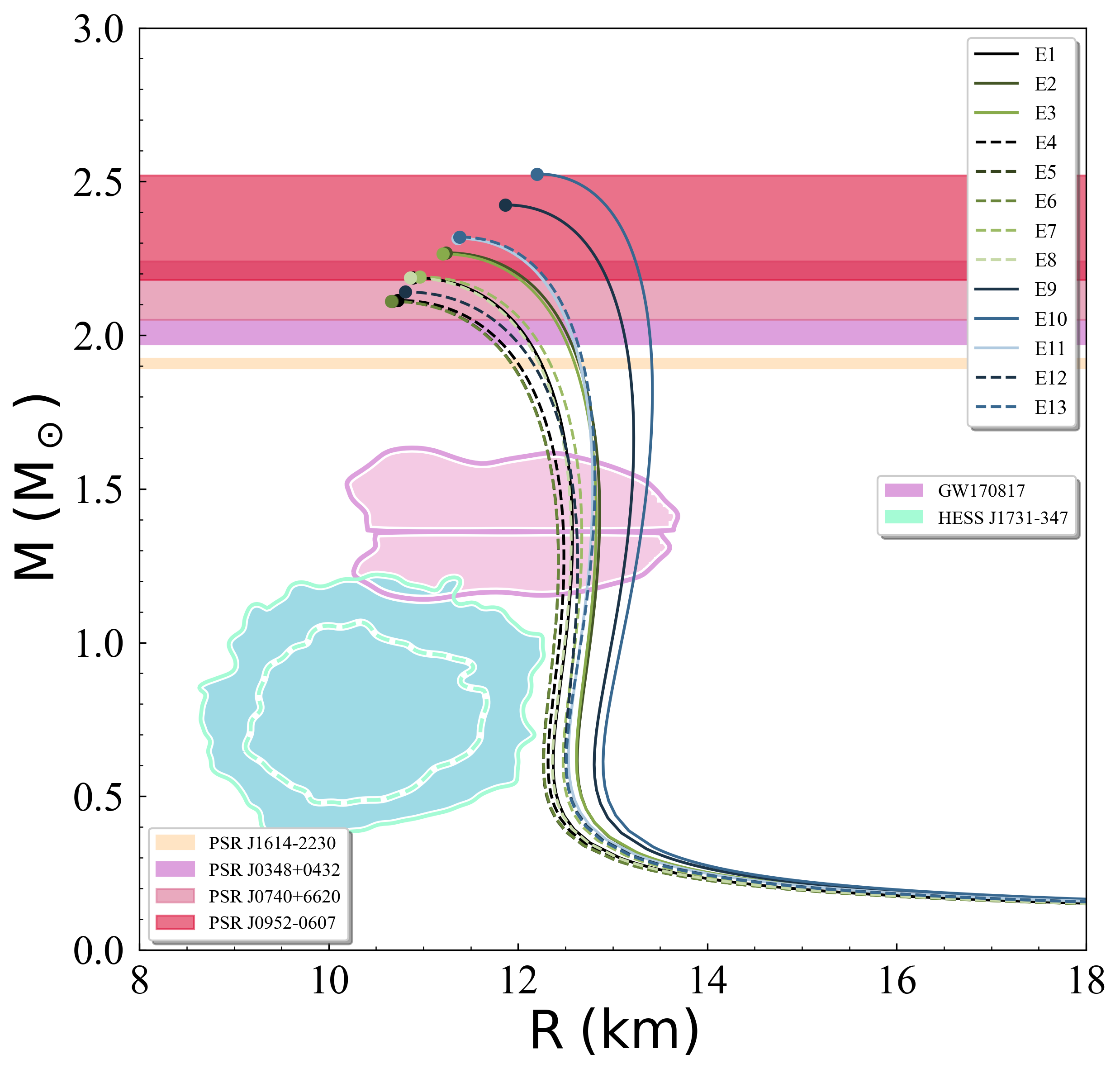}
    \caption{The M-R diagram corresponds to the RMF model of PNM for various EoSs (for more details see the text). The shaded regions from bottom
to top represent the HESS J1731-347 remnant~\cite{Doroshenko-2022}, the GW170817 event~\cite{Abbott-2019-X}, PSR J1614-2230~\cite{Arzoumanian-2018}, PSR J0348+0432~\cite{Antoniadis-2013}, PSR J0740+6620~\cite{Cromartie-2020}, 
and PSR J0952-0607~\cite{Romani-2022} pulsar observations for the possible maximum mass.}
    \label{fig:mr_1}
\end{figure}

Starting with the RMF nuclear model of Pure Neutron Matter (PNM), we see that its parameterization makes the EoSs to present two different behaviors. The general tendency is, as we can see in the Fig.\ref{fig:mr_1}, that the increase of the admixture fraction coefficient $a_X$ leads to stiffer EoS and on the other hand the increase of the $\rho$-coupling leads to a softer EoS. The combination between these two parameters can be optimized, resulting a reasonable EoS which fits inside the accepted limits with respect the observational data.

Performing this optimization we can deduce that an admixture of $30\%$ in the $a_X$ coefficient and using a $\rho$-coupling of +$5\%$ results a stiff EoS with maximum mass of $\mathrm{2.33\;M_\odot}$. Furthermore an admixture of $20\%$ in $a_X$ in combination with the $\rho$-coupling varying between +$5\%$ and +$10\%$ results EoSs with maximum masses from $\mathrm{2.1\;M_\odot}$ to $\mathrm{2.27\;M_\odot}$ and radius between 12.5 Km to 13 Km respectively. We notice that none of the total EoSs of the RMF model that we used in our work can describe the HESS J1731-347~\cite{Doroshenko-2022}.

\begin{figure}[h]
    \centering
    \includegraphics[width=80mm, height=70mm]{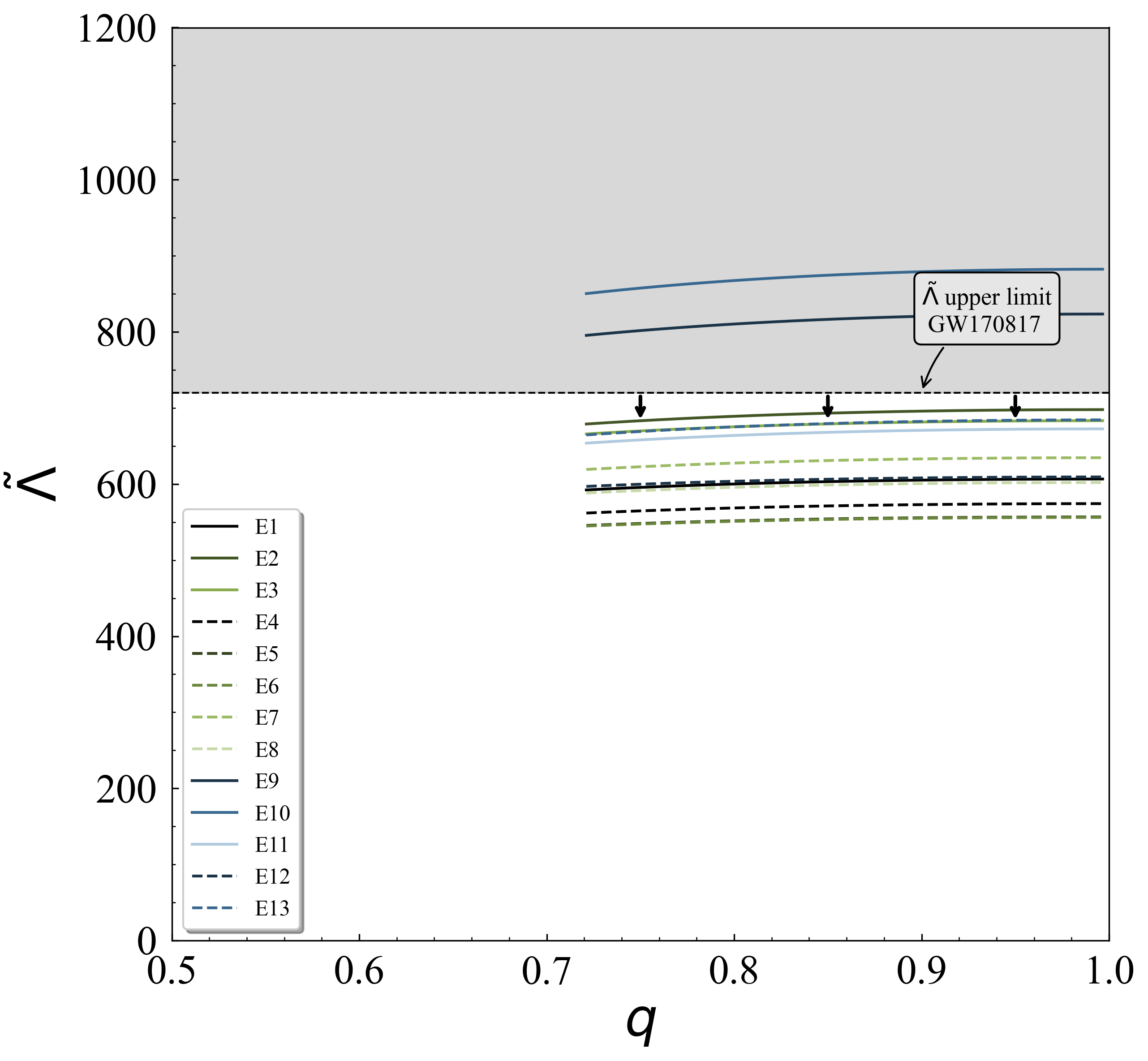}
    \caption{The $\tilde{\Lambda}-q$ dependence corresponds to the RMF model of PNM. The shaded region shows the excluded values derived from the GW170817 event~\cite{Abbott-2019-X}.  }
    \label{fig:lambda-q}
\end{figure}

In addition as we can see from Fig.\ref{fig:lambda-q}, in the ${\tilde\Lambda}-q$ diagram concerning the effective tidal deformability with respect to the binary mass ratio $q$, the aforementioned optimization can exist well below the specified upper limit of the first gravitational wave GW170817 observation given by LIGO~\cite{Abbott-2019-X}. In our study we adopted the estimations of GW170817 detection concerning the component masses and the chirp mass of the system~\cite{Abbott-2019-X}. As one can observe from Fig.\ref{fig:lambda-q}, the observational upper limit on $\tilde{\Lambda}$ favors in general the $20\%$ admixture, contrary to the $30\%$ admixture, as the first one provides softer EoSs. Across the same amount of admixture, the EoSs with higher $\rho$-coupling provide lower values of $\tilde{\Lambda}$. Though, this is not an all-out decisive behavior for the exclusion of the $30\%$ admixture since there are cases that fulfill the observational upper limit on $\tilde{\Lambda}$.

\begin{figure}[h]
    \centering
    \includegraphics[width=80mm, height=70mm]{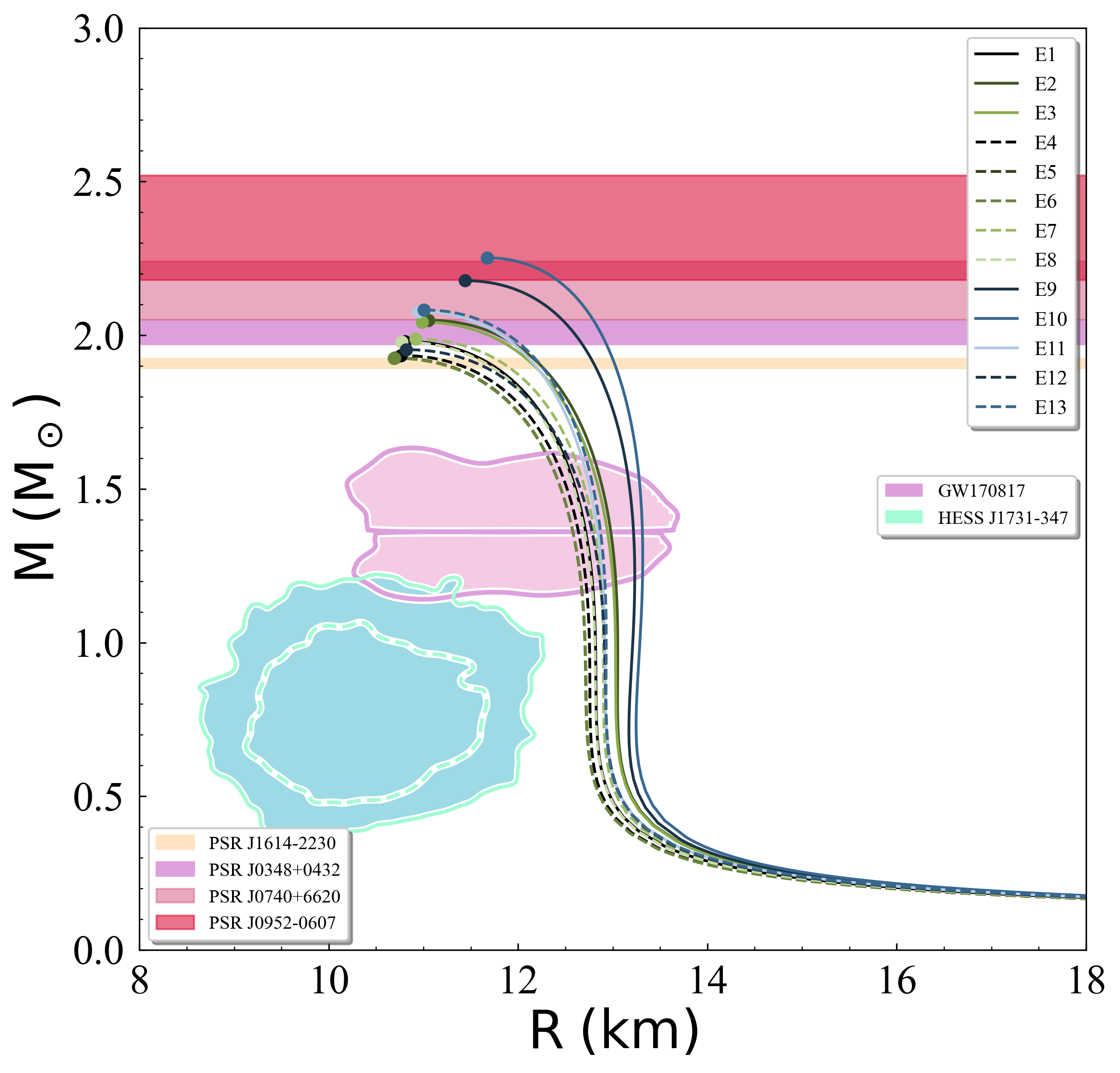}
    \caption{The M-R diagram corresponds to the RMF model of $\beta$-equilibrated (npe) matter for various EoSs (for more details see the text). The shaded regions from bottom to top represent the HESS J1731-347 remnant~\cite{Doroshenko-2022}, the GW170817 event~\cite{Abbott-2019-X}, PSR J1614-2230~\cite{Arzoumanian-2018}, PSR J0348+0432~\cite{Antoniadis-2013}, PSR J0740+6620~\cite{Cromartie-2020}, 
and PSR J0952-0607~\cite{Romani-2022} pulsar observations for the possible maximum mass.}
    \label{fig:mr_rmf_beq}
\end{figure}

\begin{figure}[h]
    \centering
    \includegraphics[width=80mm, height=70mm]{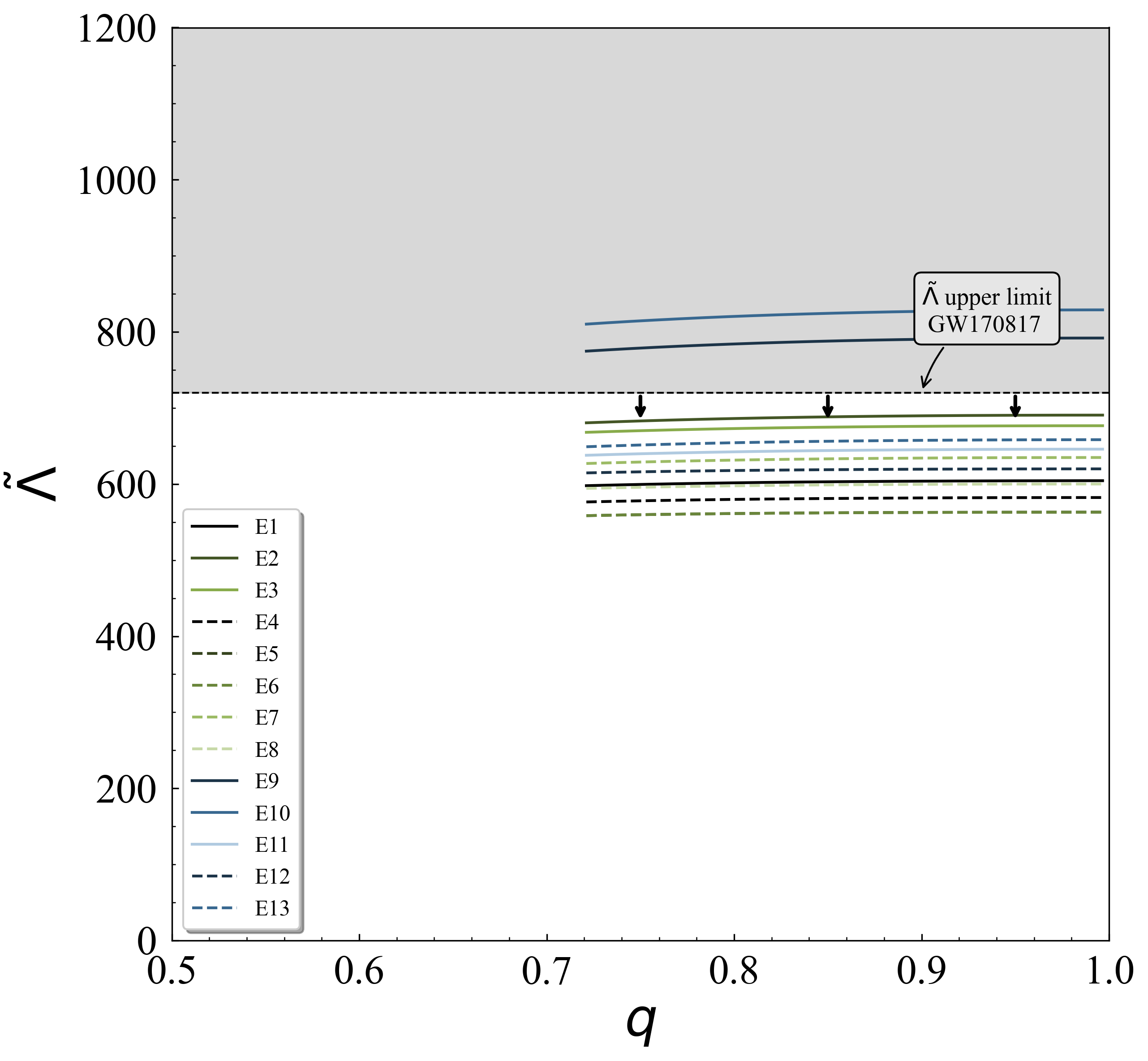}
    \caption{The $\tilde{\Lambda}-q$ dependence corresponds to the RMF model of $\beta$-equilibrated (npe) matter. The shaded region shows the excluded values derived from the GW170817 event~\cite{Abbott-2019-X}.}
    \label{fig:lambda-q-beq}
\end{figure}

When considering $\beta$-equilibrated (npe) matter instead of PNM in the RMF model and the admixture into vector boson segment constrained by 20\% and 30\% we see not much of a difference of neutron stars radius. A appreciable reduction appears in the maximum mass as shown in Fig.\ref{fig:mr_rmf_beq}.
Looking to the relative $\tilde{\Lambda}$ we see a good agreement with the constraints from GW170817 observational data as depicted in Fig.\ref{fig:lambda-q-beq}. 

Furthermore, we studied the effect of the X17 boson using the MDI model. To be more specific, we used as a base three main MDI EoSs with different slope parameter ${\mathrm{L}}$: $\mathrm{L=65\;MeV}$, $\mathrm{L=72.5\;MeV}$, and $\mathrm{L=80\;MeV}$, with blue, green, and red color in the corresponding figures respectively. The coupling constant $\mathrm{g}$ is allowed to take values up to $\mathrm{g\simeq0.022}$ to fulfill the properties of symmetric nuclear matter at $\mathrm{n_0}$, while the parameter $\mathrm{C}$ ranges in the interval $\mathrm{C\in[0,C_{max}]}$, where $\mathrm{C_{max}}$ is the higher allowed value, unique for each EoS, derived from the non-violation of causality. To be more specific, the value of $\mathrm{g_{max}\simeq0.022}$ arises from the deviation of the binding energy of the symmetric nuclear matter at the saturation density, i.e. $\mathrm{-17\;Mev\leq E_{bind}(n_0)\leq-15\;MeV}$.

\begin{figure}[t]
    \centering
    \includegraphics[width=80mm, height=70mm]{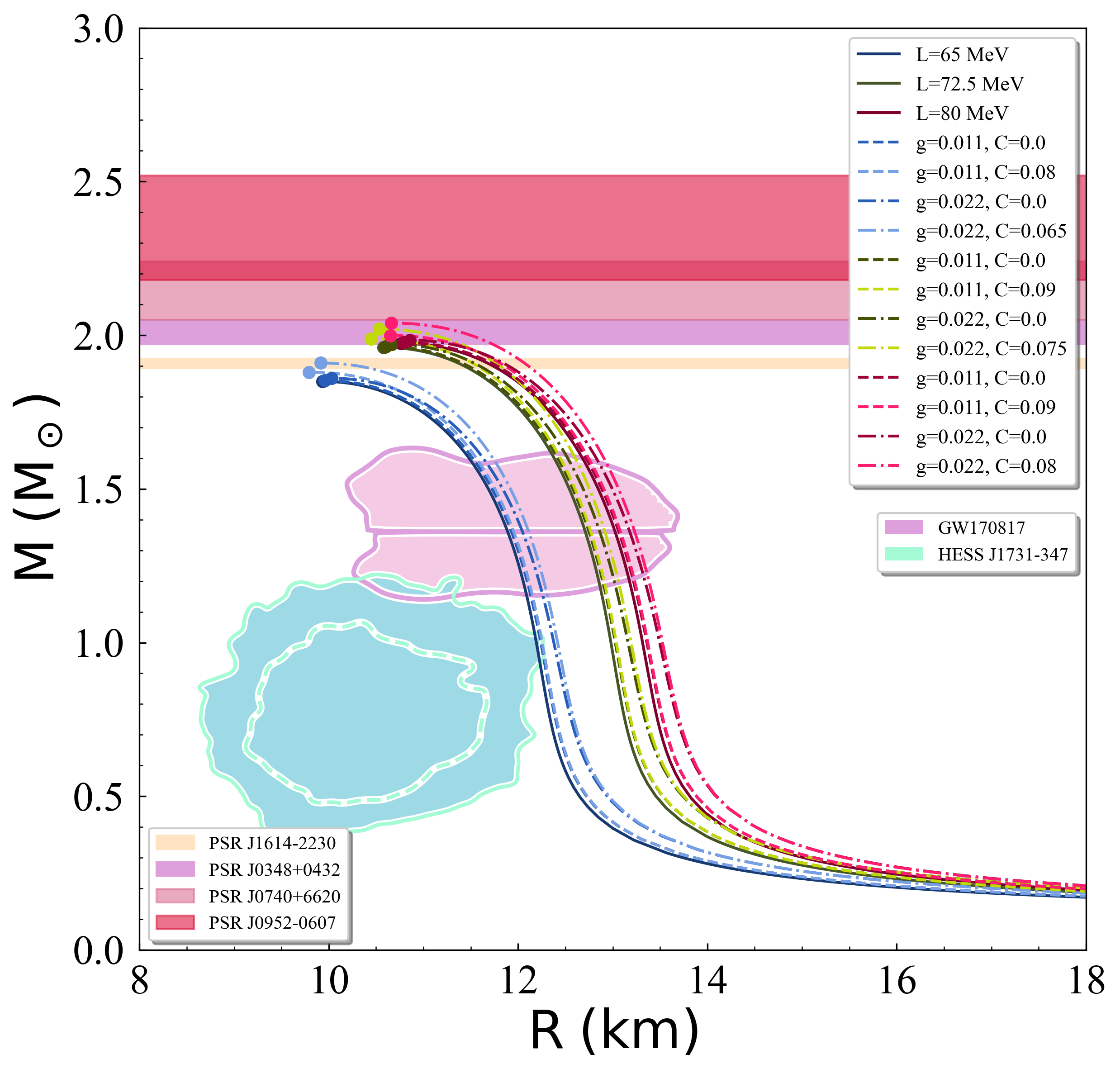}
    \caption{The M-R diagram corresponds to the MDI models. The shaded regions from bottom
to top represent the HESS J1731-347 remnant~\cite{Doroshenko-2022}, the GW170817 event~\cite{Abbott-2019-X}, PSR J1614-2230~\cite{Arzoumanian-2018}, PSR J0348+0432~\cite{Antoniadis-2013}, PSR J0740+6620~\cite{Cromartie-2020}, 
and PSR J0952-0607~\cite{Romani-2022} pulsar observations for the possible maximum mass.}
    \label{fig:mr_1-mdi}
\end{figure}

In Fig,~\ref{fig:mr_1-mdi} the mass-radius dependence for all the cases of the MDI model is displayed. The solid curves correspond to the three initial EoSs without the contribution of the X17 boson, while the dashed and dash-dotted curves correspond to the EoSs within the presence of the X17 boson for $\mathrm{g=0.011}$, and $\mathrm{g=0.022}$ respectively. Among the same set of EoS and $\mathrm{g}$, the darker colored curve corresponds to $\mathrm{C=0}$, while the lighter one to $\mathrm{C=C_{max}}$. At first sight, the higher value of $\mathrm{L}$ provides stiffer EoS, nevertheless the amount of contribution of X17. The EoSs with $\mathrm{g=0.011}$ (dashed curves) lie closely to their corresponding initial EoS (solid curves) leading to a slight increment of $\mathrm{M_{max}}$ and radius $\mathrm{R}$, while the bigger differentiation from the initial EoSs is for $\mathrm{g=0.022}$. Also, the effect from the contribution of the X17 is stronger on the coupling constant $\mathrm{g}$ than the parameter $\mathrm{C}$. We notice that all the EoSs lie outside the estimated region for the HESS J1731-347 remnant~\cite{Doroshenko-2022}.


\begin{figure}[h]
    \centering
    \includegraphics[width=80mm, height=70mm]{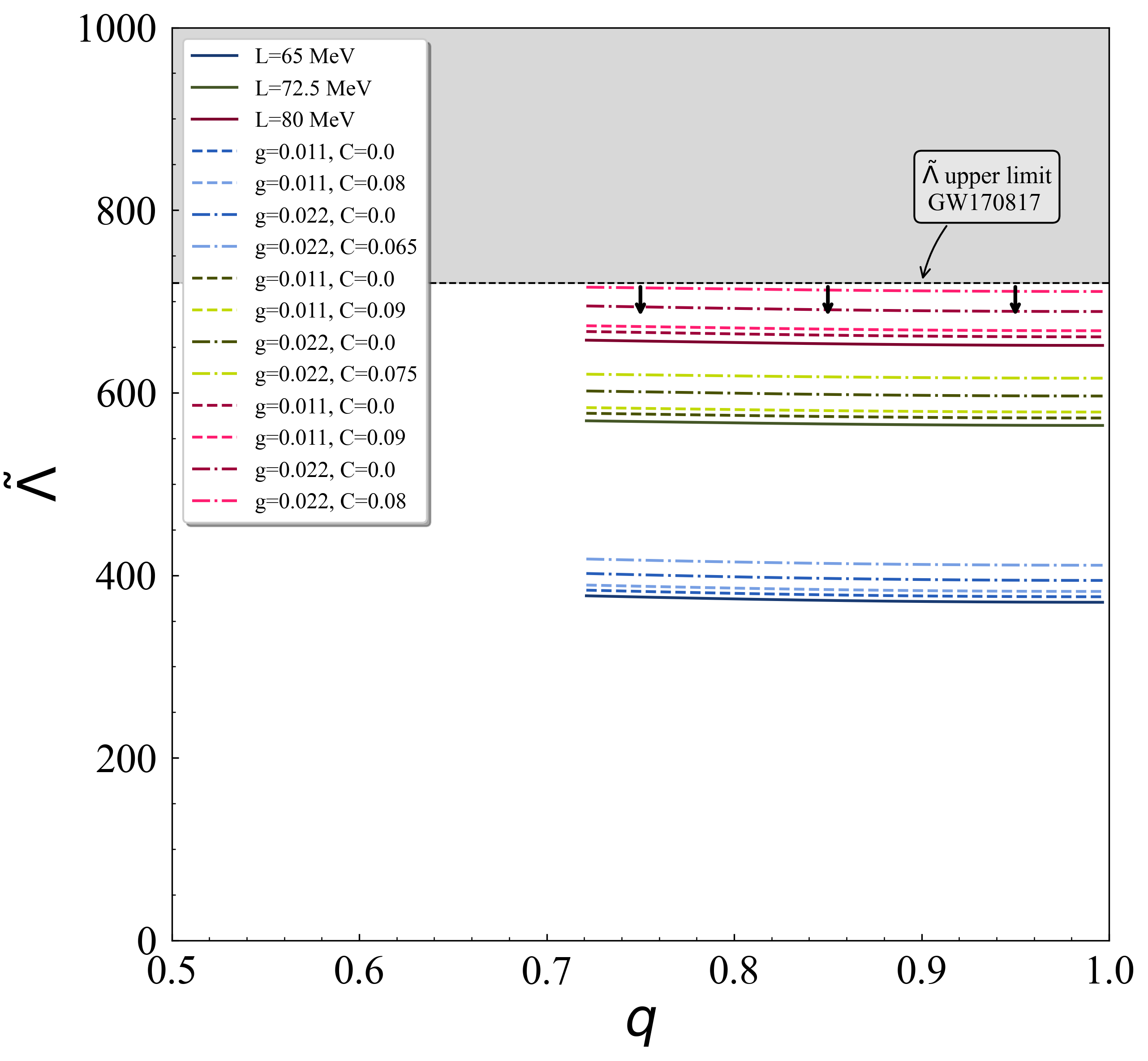}
    \caption{The $\tilde{\Lambda}-q$ dependence corresponds to the MDI models. The shaded region shows the excluded values derived from the GW170817 event~\cite{Abbott-2019-X}. } 
    \label{fig:lambda-q-mdi}
\end{figure}

\begin{figure*}[b]
\centering
    \includegraphics[width=80mm, height=70mm]{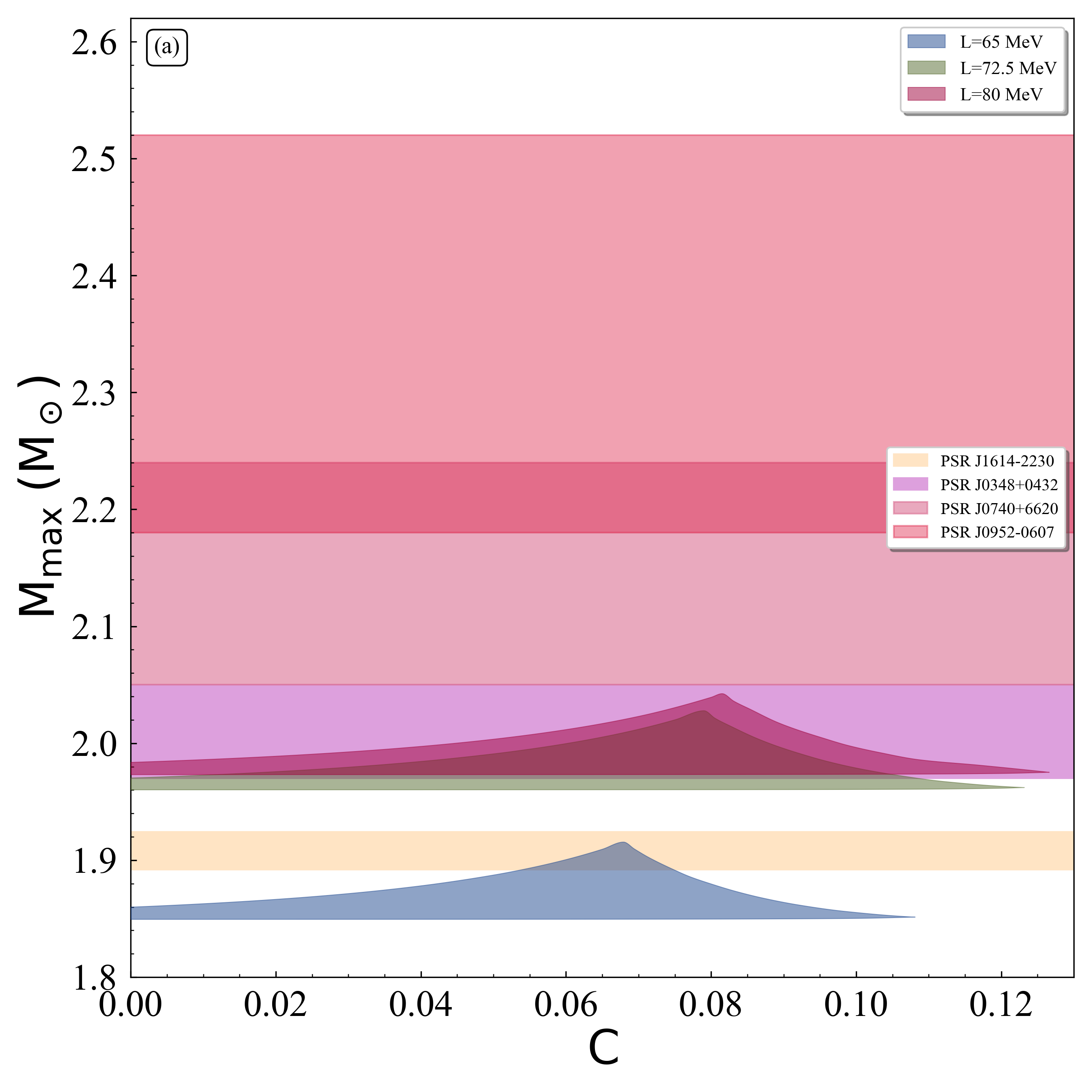}
    ~\includegraphics[width=80mm, height=70mm]{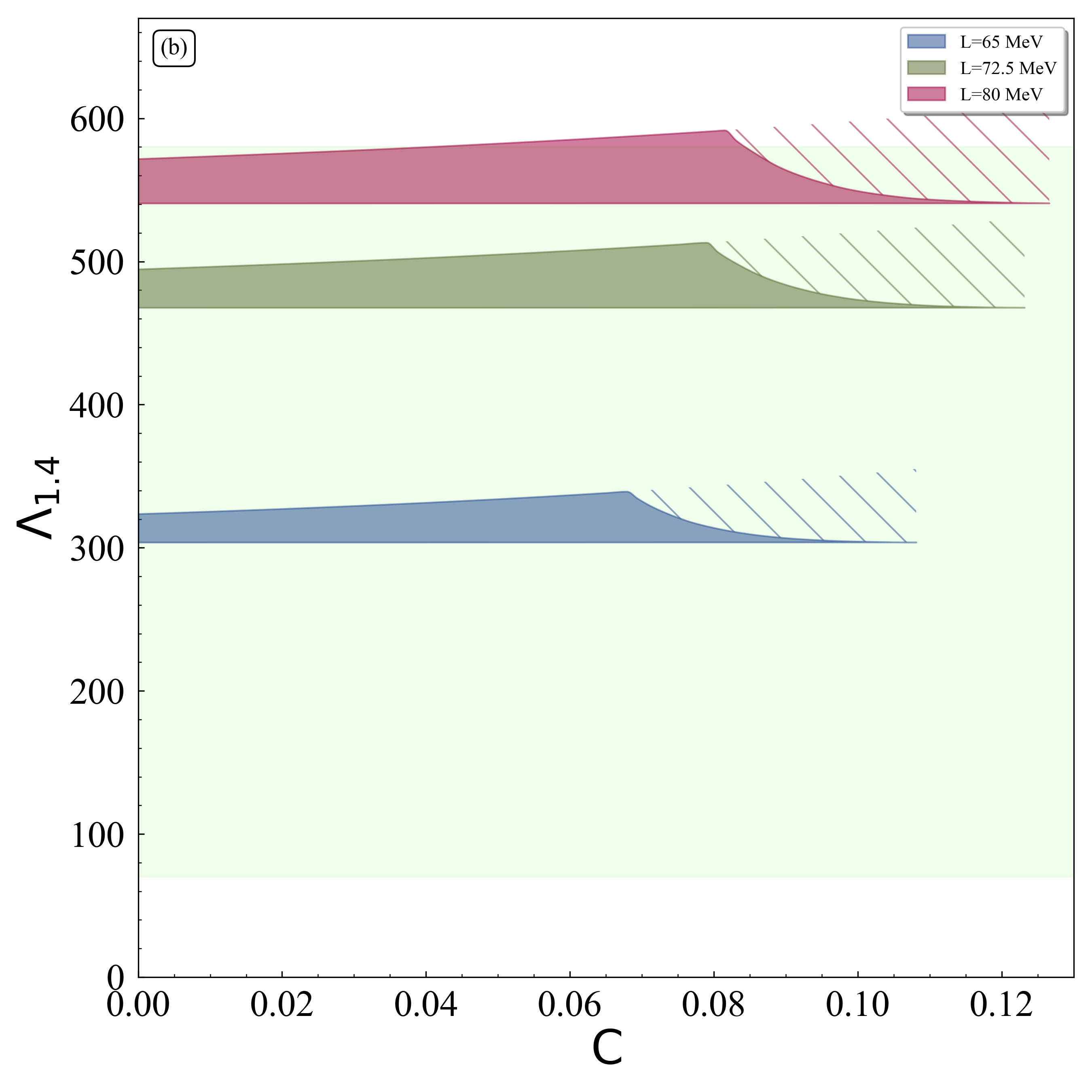}
    \caption{(a) The maximum mass, and (b) the tidal deformability $\Lambda_{1.4}$ of a $\mathrm{1.4\;M_\odot}$ neutron star related to the parameter C for the three MDI EoSs. The horizontal shaded areas on the left panel correspond to those of Figure 1. The green shaded area on the right panel indicates the constraints from GW170817~\cite{Abbott-2018}, while the colored diagonal lines show the excluded regions from the violation of causality for each EoS.}   
    \label{fig:Mmax-L14-C-mdi}
\end{figure*} 

The expansion of the aforementioned EoSs by their application to the observational data of the GW170817 event, allowed us to study directly their behavior through the $\tilde{\Lambda}-q$ dependence, as one can observe in Fig.~\ref{fig:lambda-q-mdi}. In this figure, the first two parametrizations for $\mathrm{L}$ (blue and green curves) are in a first instance in good agreement with the observational upper limit on $\tilde{\Lambda}$. The third and more stiff branch of EoSs (red curves) also lies inside the acceptance area (imposed by the upper limit of $\tilde{\Lambda}$), but it us much closer to the limit, with the more stiff EoS of the total MDI EoSs to be almost exact on it. Additionally, as we increase the slope parameter $\mathrm{L}$, the EoSs spread out more, e.g. for $\mathrm{L=65\;MeV}$ the effect of the X17 is smaller compared to the other EoSs in total. At a second level, across EoSs with the same $\mathrm{L}$, this effect depends more on the coupling constant $\mathrm{g}$, as we already observed from Fig.~\ref{fig:mr_1-mdi}. Therefore, a more detailed examination of the dependence of the EoSs on the two main parameters $\mathrm{g}$, and $\mathrm{C}$ is needed.

In Figure~\ref{fig:Mmax-L14-C-mdi}, we examined the dependence of the EoSs from the $\mathrm{C}$ parameter using observational constraints on the $\mathrm{M_{max}}$ and $\mathrm{\Lambda_{1.4}}$. By focusing on the left panel of Figure~\ref{fig:Mmax-L14-C-mdi}, we notice that at the most high contribution of the X17, the maximum mass increased $\mathrm{\Delta M_{max}\simeq0.066\;M_\odot}$ for $\mathrm{L=65\;MeV}$, $\mathrm{\Delta M_{max}\simeq0.068\;M_\odot}$ for $\mathrm{L=72.5\;MeV}$, and $\mathrm{\Delta M_{max}\simeq0.069\;M_\odot}$ for $\mathrm{L=80\;MeV}$. However, for the same value of $(\mathrm{g,C})$ bewteen the three different set of MDI EoSs, the $\mathrm{\Delta M}$ is higher for the set with the lower $\mathrm{L}$; e.g. for $\mathrm{g=0.022}$ and $\mathrm{C=0.06}$ the corresponding values are $\mathrm{\Delta M_{max}^{L=65}\simeq0.06\;M_\odot}$, $\mathrm{\Delta M_{max}^{L=72.5}\simeq0.045\;M_\odot}$, and $\mathrm{\Delta M_{max}^{L=80}\simeq0.043\;M_\odot}$. Hence, the effect of the X17 under this perspective is as a relative amount higher in the softer EoSs. Moreover, the EoS with $\mathrm{L=65\;MeV}$ fails to provide a neutron star with $\mathrm{M_{max}\geq2.0\;M_\odot}$ and requires high value of the coupling constant $\mathrm{g}$ in order to provide a neutron star with $\mathrm{M_{max}\simeq1.9\;M_\odot}$, such as the PSR J1614-2230~\cite{Arzoumanian-2018}. In the other two cases of $\mathrm{L}$, the presence of the X17 could lead them to predict a $\mathrm{\geq2.0\;M_\odot}$ neutron star (for a variety of combinations of $\mathrm{g}$ and $\mathrm{C}$) but not higher than the $\mathrm{\sim2.05\;M_\odot}$ value. The "shark-fin" shaded region arises from the constraints that the non-violation of the causality implies on the $\mathrm{C_{max}}$, with the peak corresponding to the $\mathrm{(g=0.022,C=C_{max})}$ pair of values for each one of the three MDI set of EoSs. In addition the shaded regions for $\mathrm{L=72.5\;MeV}$ and $\mathrm{L=80\;MeV}$ are hardly be distinguished. 

By examining further the way in which the contribution dependence of the X17 and the corresponding EoSs on the $\mathrm{C}$ parameter evolves, we constructed the Fig.~\ref{fig:Mmax-L14-C-mdi}(b), in which the behavior of the dimensionless tidal deformability $\Lambda_{1.4}$ of a $\mathrm{1.4\;M_\odot}$ neutron star related to the $\mathrm{C}$ for all the MDI EoSs is shown. The shape of the shaded regions is similar to that of the left panel of Fig.~\ref{fig:Mmax-L14-C-mdi} and each peak corresponds to the $\mathrm{(g=0.022,C=C_{max})}$ combination, unique for each set of EoSs. In both panels, as we move to higher values of $\mathrm{L}$ the peak, meaning the higher contribution of X17 for each EoS, is shifted to higher values of $\mathrm{C}$. The diagonal colored lines indicate the area of values of $\mathrm{g}$ and $\mathrm{C}$ that violates the causality.

Contrary to the behavior that we observed in the $\mathrm{M_{max}-C}$ diagram, the $\mathrm{L=72.5\;MeV}$ and $\mathrm{L=80\;MeV}$ families of EoSs are clearly distinguished. Especially, the $\mathrm{L=80\;MeV}$ set of EoSs lies almost inside the observational data from LIGO (green shaded area) with a small peak violating the observational upper limit of $\Lambda_{1.4}$. The other two families of MDI EoSs lie inside the observational data for all the combinations of $\mathrm{g}$ and $\mathrm{C}$ that we used. Therefore, even if the $\mathrm{L=72.5\;MeV}$ and $\mathrm{L=80\;MeV}$ set of EoSs provided almost identical maximum masses as we showed in Fig.~\ref{fig:Mmax-L14-C-mdi}(a), their dependence on $\Lambda_{1.4}$ (i.e. the radius $\mathrm{R_{1.4}}$) highlights their differences, and especially the sensitivity of the radius to the X17 contribution. Another issue that arises is the tension between the favor of softer EoS from the tidal deformability perspective, and the requirement for a stiffer EoS in order to provide a sufficient high maximum neutron star mass. This tension, through the corresponding observations, we used as a tool in our study to impose constraints on the contribution of X17.

\begin{figure}[h]
    \centering
    \includegraphics[width=80mm, height=70mm]{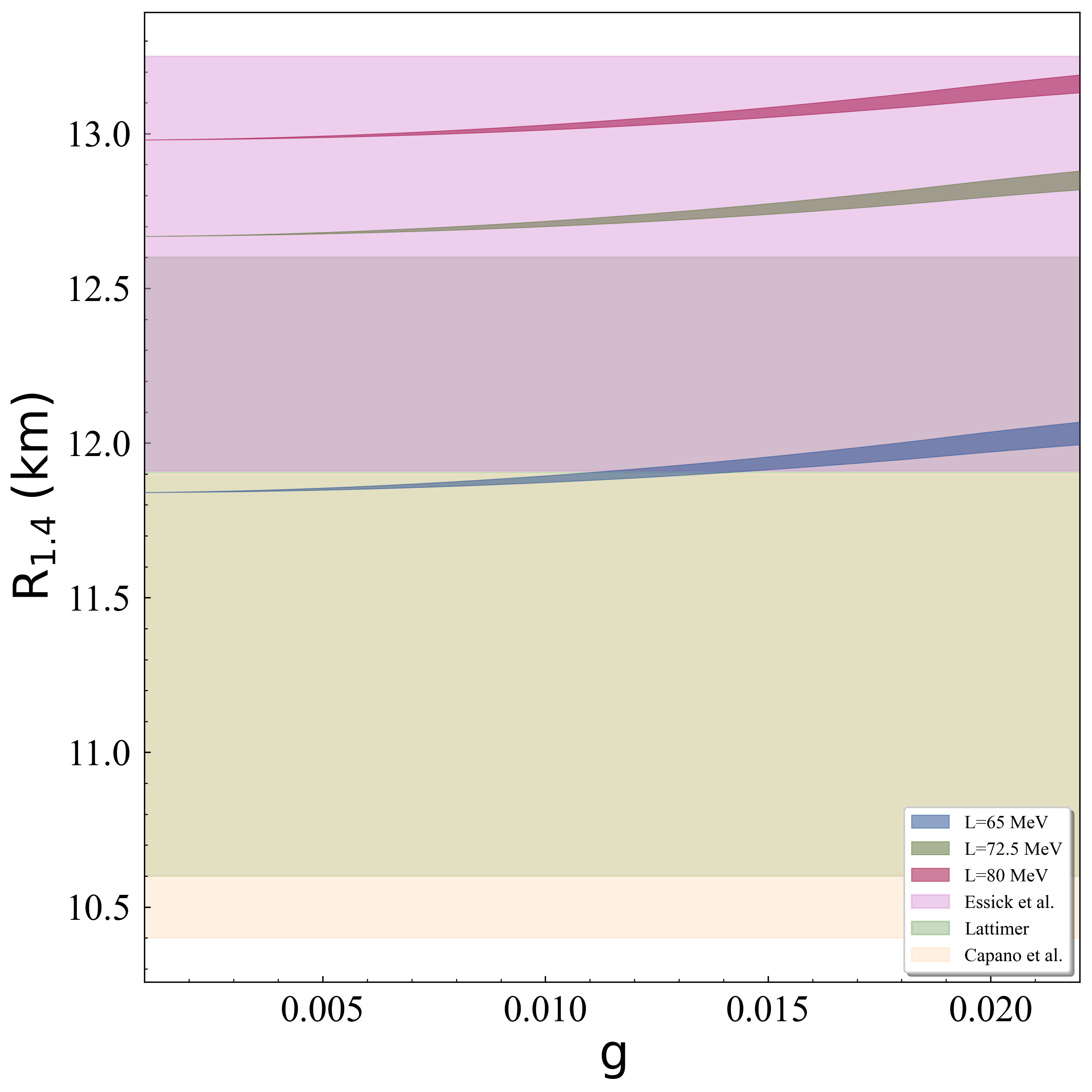}
    \caption{The radius $\mathrm{R_{1.4}}$ of a $\mathrm{1.4\;M_\odot}$ neutron star related to the coupling constant $\mathrm{g}$ for all the MDI EoSs that we used in our study. The horizontal shaded regions correspond to different constraints on $\mathrm{R_{1.4}}$~\cite{Lattimer-2023,Capano-2020,Essick-2020}.}
    \label{fig:R14-g-MDI}
\end{figure}

As we mentioned before, the study of the X17 related to radius $\mathrm{R_{1.4}}$ is of interest. In Fig.~\ref{fig:R14-g-MDI} we show the behavior of $\mathrm{R_{1.4}}$ with respect to the coupling constant $\mathrm{g}$ for all the three sets of MDI EoSs. The horizontal shaded regions indicate various estimations of $\mathrm{R_{1.4}}$~\cite{Lattimer-2023,Capano-2020,Essick-2020}. The $\mathrm{L=72.5\;MeV}$ and $\mathrm{L=80\;MeV}$ EoSs lie inside only on the estimated area of Ref.~\cite{Essick-2020}. On the other hand, the EoSs with $\mathrm{L=65\;MeV}$ lie inside the values predicted by Refs.~\cite{Lattimer-2023,Capano-2020}, and above the coupling constant $\mathrm{g\simeq0.011}$ the corresponding EoSs insert the region of Ref.~\cite{Essick-2020}, but violate the upper limit of Ref.~\cite{Capano-2020}. As a general remark, as we move to higher values of $\mathrm{g}$ the effect of the X17 boson becomes bigger, leading to bigger radii. Also, for higher values of $\mathrm{g}$ the dependence from the $\mathrm{C}$ parameter is stronger (thickening of the curved shaded region), contrary to the lower values of $\mathrm{g}$ where the impact of different $\mathrm{C}$ on the radius $\mathrm{R_{1.4}}$ vanishes.

\begin{figure}
    \centering
    \includegraphics[width=80mm, height=70mm]{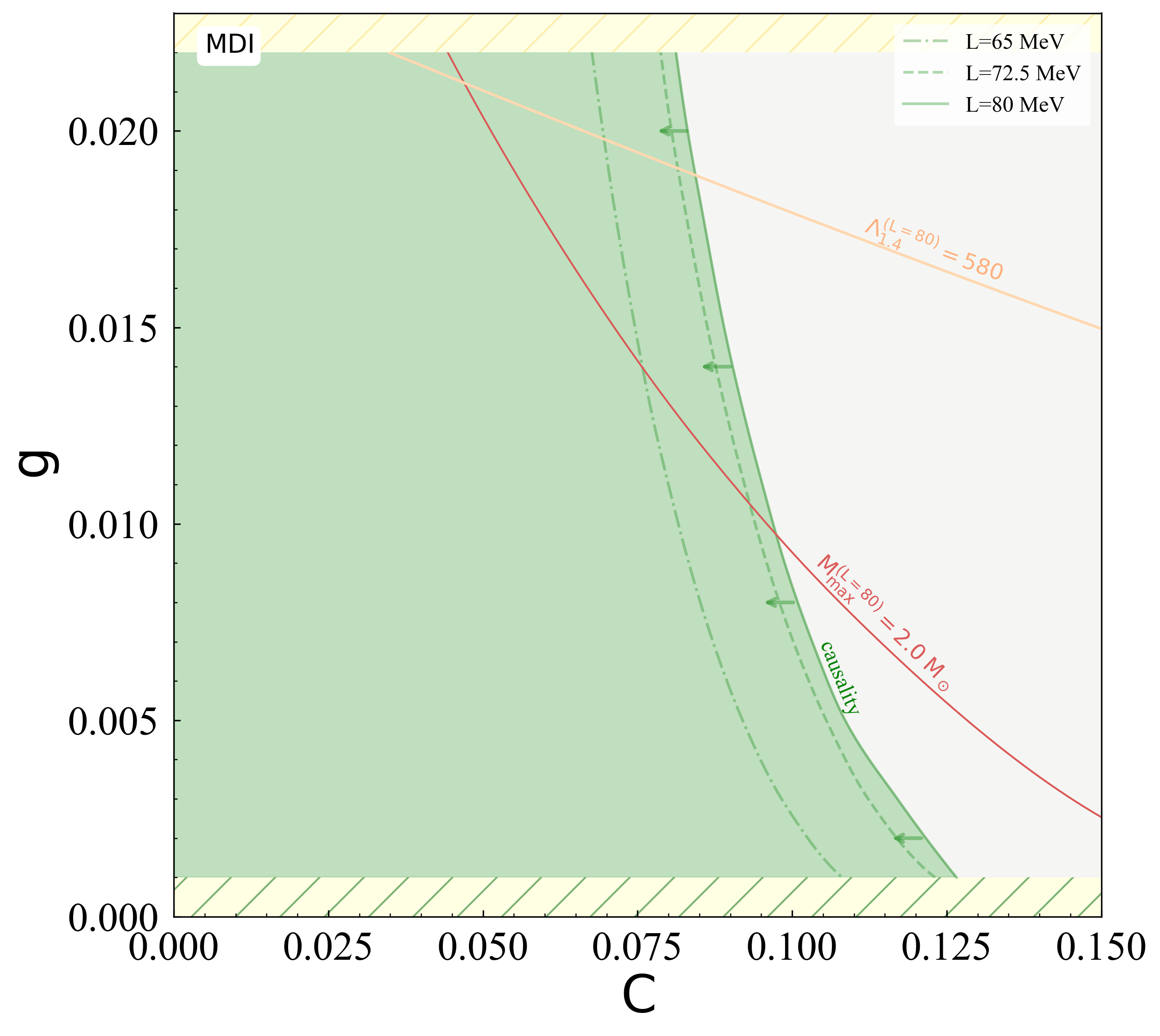}
    \caption{Constraints for $\mathrm{g}$ and $\mathrm{C}$ for the three MDI EoSs with respect to causality, the possible upper mass limit $\mathrm{M_{max}=2\;M_{\odot}}$ and the dimensionless tidal $\Lambda_{1.4}=580$ for the MDI (L=80 MeV) EoS.}
    \label{g-C-MDI}
\end{figure}

So far we studied the two main parameters, $\mathrm{g}$ and $\mathrm{C}$, through their dependence on the macroscopic properties of neutron stars, derived from the EoS, and by exploiting the available observations. In general, only for higher coupling constant $\mathrm{g}$ ($\mathrm{g\gtrapprox0.009}$) the EoS becomes more sensitive to the $\mathrm{C}$. In order to examine further how $\mathrm{g}$ and $\mathrm{C}$ affects the EoSs, we introduced a $\mathrm{g-C}$ parameter space, displayed in Fig.~\ref{g-C-MDI}. This figure is suitable for an overall view on these parameters and the way that the constraints affect them. The green shaded region shows the allowed parameter space with respect to causality, for all three set of MDI EoSs. The green arrows show the direction of the accepted region. The green solid curve indicates the limit for the $\mathrm{L=80\;MeV}$ EoS, the dashed one indicates the limit for the $\mathrm{L=72.5\;MeV}$ EoS, while the dash-dotted curve indicates the limit for the $\mathrm{L=65\;MeV}$ EoS. 

As one can observe, the constraint imposed from the non-violation of causality, cuts off the very high values of $\mathrm{g}$ and $\mathrm{C}$ (grey shaded area). Additionally, as we move to EoSs with lower $\mathrm{L}$, this bound is shifted to even lower values of the parameter space, minimizing further the "window". The red solid curve shows the maximum mass limit of $\mathrm{M_{max}=2\;M_\odot}$ for the $\mathrm{L=80\;MeV}$ EoS; for the parameter space on the left of this curve, the combination of $(\mathrm{g,C})$ leads to neutron stars with $\mathrm{M_{max}<2\;M_\odot}$. The light orange curve, shows the observational upper limit of $\mathrm{\Lambda_{1.4}}$ for the $\mathrm{L=80\;MeV}$ EoS; only the combination of parameters on the left side of this curve can be accepted, without violating the observational value. Therefore, for the $\mathrm{L=80\;MeV}$ EoS, if we looking for a neutron star with at least $\mathrm{M_{max}\simeq2\;M_\odot}$ with respect to $\mathrm{\Lambda_{1.4}}$ and causality, then we must search inside the curved triangular parameter space that the three curves form. We notice that this space is unique for each set of EoS, and the corresponding $\mathrm{M_{max}}$ requirement. Summarizing, strong constraints are introduced both by observational data and causality leading to a significant limitation of the allowed range of parameters; a behavior that is present correspondingly in the case of QSs, as we will demonstrate below.

\begin{figure}
    \centering
    \includegraphics[width=80mm, height=70mm]{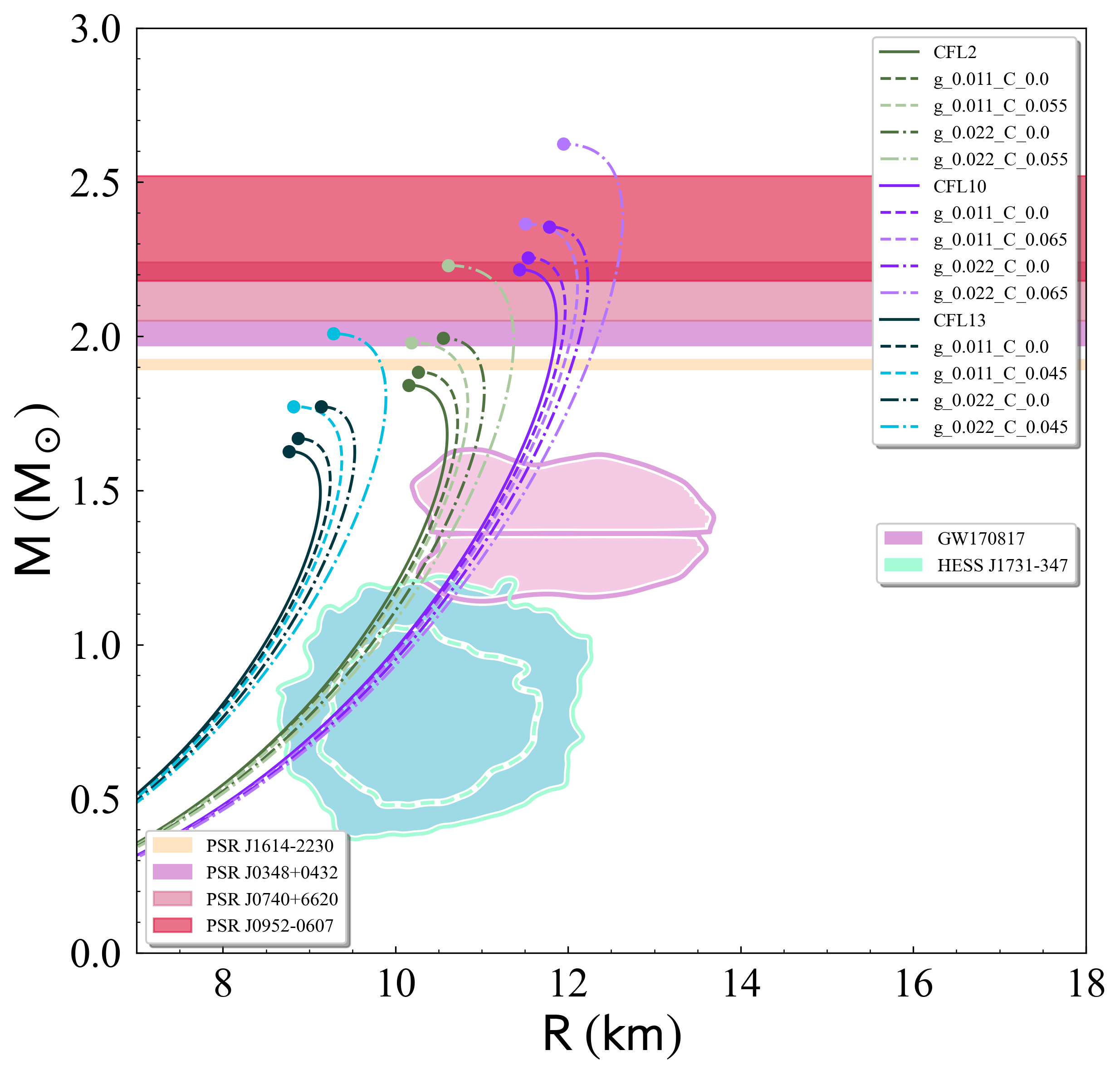}
    \caption{The M-R diagram of the three CFL quark EoSs. The shaded regions from bottom to top represent the HESS J1731-347 remnant~\cite{Doroshenko-2022}, the GW170817 event~\cite{Abbott-2019-X}, PSR J1614-2230~\cite{Arzoumanian-2018}, PSR J0348+0432~\cite{Antoniadis-2013}, PSR J0740+6620~\cite{Cromartie-2020}, and PSR J0952-0607~\cite{Romani-2022} pulsar observations.}
    \label{fig:mr_cfl}
\end{figure}

\begin{figure}[h]
    \centering
    \includegraphics[width=80mm, height=70mm]{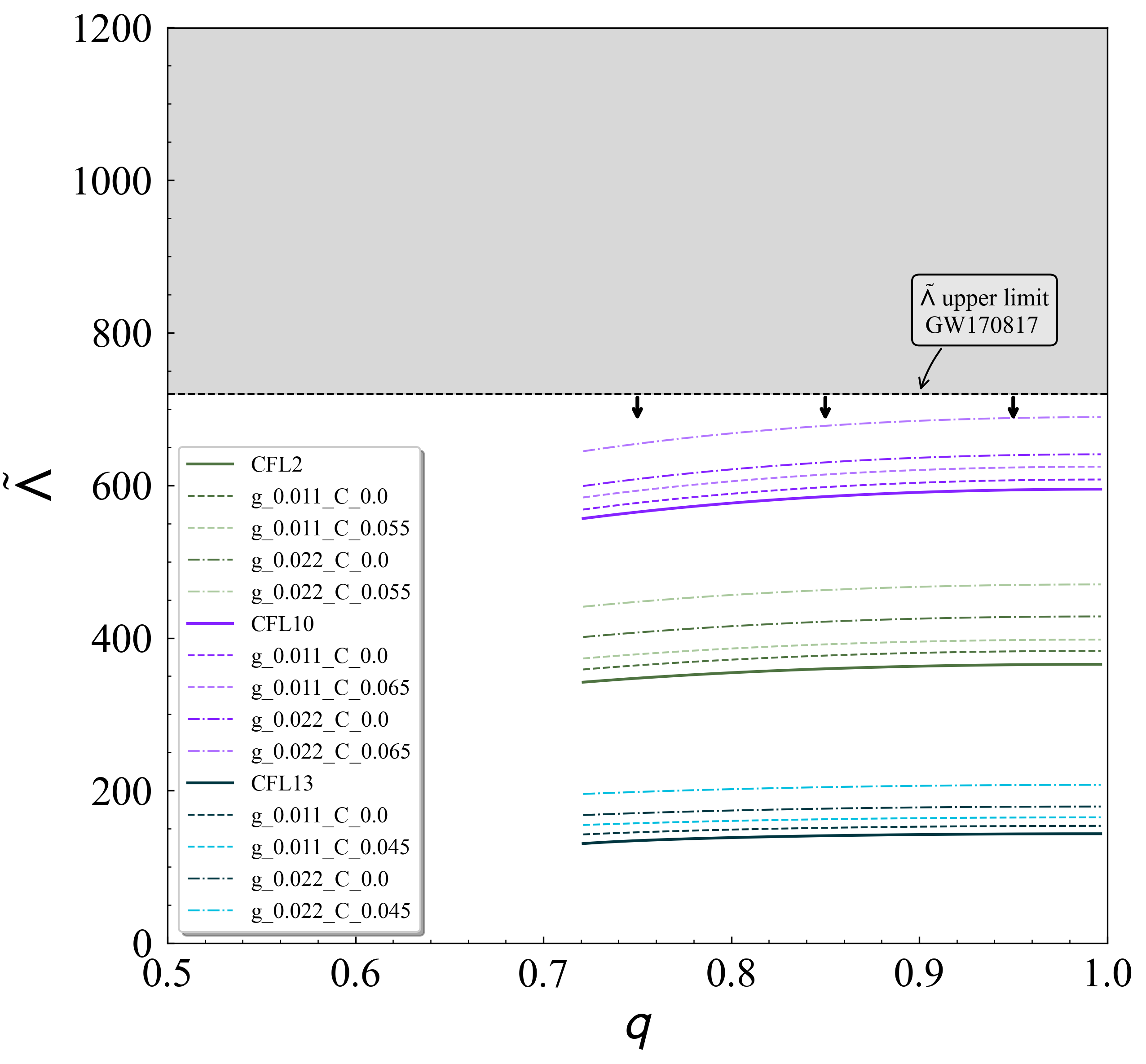}
    \caption{The dimensionless tidal defomability $\tilde{\Lambda}$ for the three quark EoSs and various parametrization for the X17 boson.  The shaded region shows the acceptance values derived from the GW170817 event~\cite{Abbott-2019-X}.}
    \label{fig:Ltilde-q-cfl}
\end{figure}

Beyond the hadronic EoSs, we expanded our study of the contribution of the X17 boson to QSs. The effect of this contribution is depicted on the mass-radius dependence in Fig.~\ref{fig:mr_cfl}. In particular, we used three different -concerning the parametrization- sets of EoSs running from soft (CFL13 EoS) and medium (CFL2 EoS) to stiff (CFL10 EoS) behavior, covering the most possible cases. We notice that in the case of QSs we chose the same upper value for $\mathrm{g}$ so that the comparison with the hadronic EoSs of the MDI model to be more clear (the quark EoSs does not have to respect the properties of symmetric nuclear matter at the saturation density). The dashed curves correspond to coupling constant $\mathrm{g}=0.011$ while the dash-dotted ones correspond to $\mathrm{g}=0.022$. For the same value of $\mathrm{g}$ the lighter colored curves correspond to higher $\mathrm{C}$. As we underlined in the hadronic EoSs, the contribution of X17 in general depends stronger on $\mathrm{g}$ than on $\mathrm{C}$. It is worth noticing that in the case of a pure QS the effects of X17 are more pronounced compared to the neutron star case, beucase of the additional contribution of factor of 9 on energy density and pressure (see Eq.~\ref{En-pre-1-quark}). Especially, the possible existence of X17 affects dramatically the provided maximum mass in each set of EoSs. The effect on the radius of a $\mathrm{1.4\;M_\odot}$ star is moderate but enough to affect the tidal deformability (see Fig.~\ref{fig:Ltilde-q-cfl}). Also, it is interesting and noteworthy that the CFL10 (stiff) and CFL13 (medium) EoSs satisfy simultaneously both the observational data and the corresponding constraints.

\begin{figure}[hb] 
    \centering
    \includegraphics[width=80mm, height=70mm]{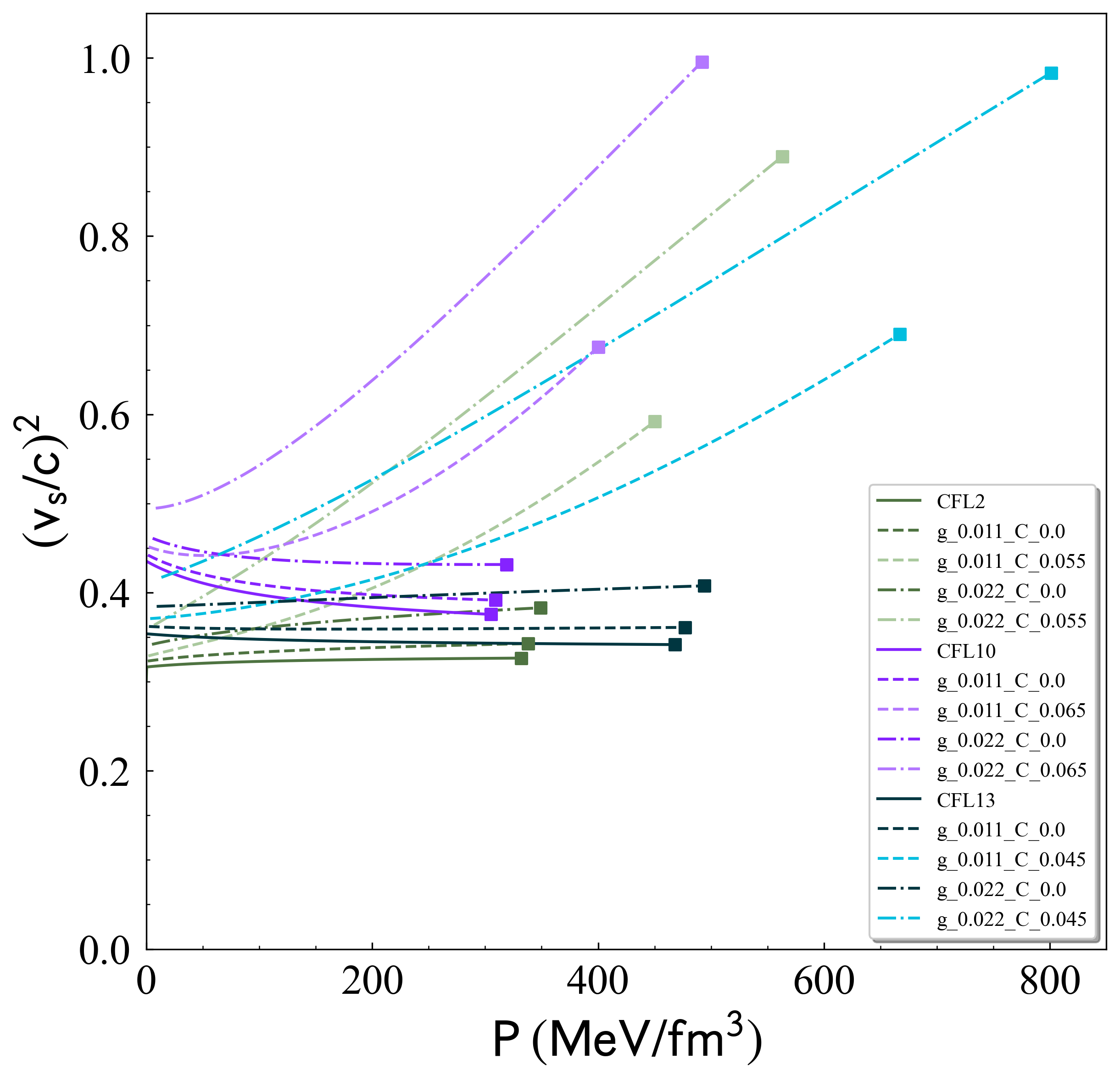}
    \caption{The speed of sound as a function of the pressure for the three quark EoSs and various parametrization for the X17 boson. The square points indicate the pair of values which provide the $\mathrm{M_{max}}$ for each CFL EoS.}
    \label{fig:speedofsound-quark}
\end{figure}

An important microscopic quantity to study the properties of the EoS is the speed of sound (see Eq.~\ref{speed-1}). In the context of this study we used so far the upper bound on the speed of sound, derived from the causality; the speed of sound should never exceed the speed of light $\mathrm{(v_s/c)^2\leq1}$. In Fig.~\ref{fig:speedofsound-quark} we demonstrate the relation between $\mathrm{(v_s/c)^2}$ and the pressure $\mathrm{P}$ for all the CFL EoSs. The square points indicate the pair of values which provide the $\mathrm{M_{max}}$ for each CFL EoS. All EoSs respect the upper bound of causality. We notice the high dependence of the speed of sound on the value of coupling constant $\mathrm{g}$. Additionally, the dependence on $\mathrm{C}$ can be observed; this dependence becomes stronger for higher values of $\mathrm{g}$.

By applying the study of quark EoSs to the observational estimations of the properties of the GW170817 event~\cite{Abbott-2019-X}, we examined further the behavior of X17 in the CFL EoSs, as shown in Fig.~\ref{fig:Ltilde-q-cfl}, through the $\mathrm{\tilde{\Lambda}-q}$ dependence. As one can observe, the effect of the X17 depends on the parametrization of both the coupling constant $\mathrm{g}$ and the  in-medium effects described by the parameter $\mathrm{C}$, leading to an appreciably increase of $\tilde{\Lambda}$. As in the case of hadronic EoSs, the dependence on $\mathrm{g}$ is stronger than on $\mathrm{C}$. In general, the stiffness on the EoS magnifies gradually this deviation on $\tilde{\Lambda}$ from the initial (without X17) EoS.

\begin{figure*}[!t]
\centering
    \includegraphics[width=80mm, height=70mm]{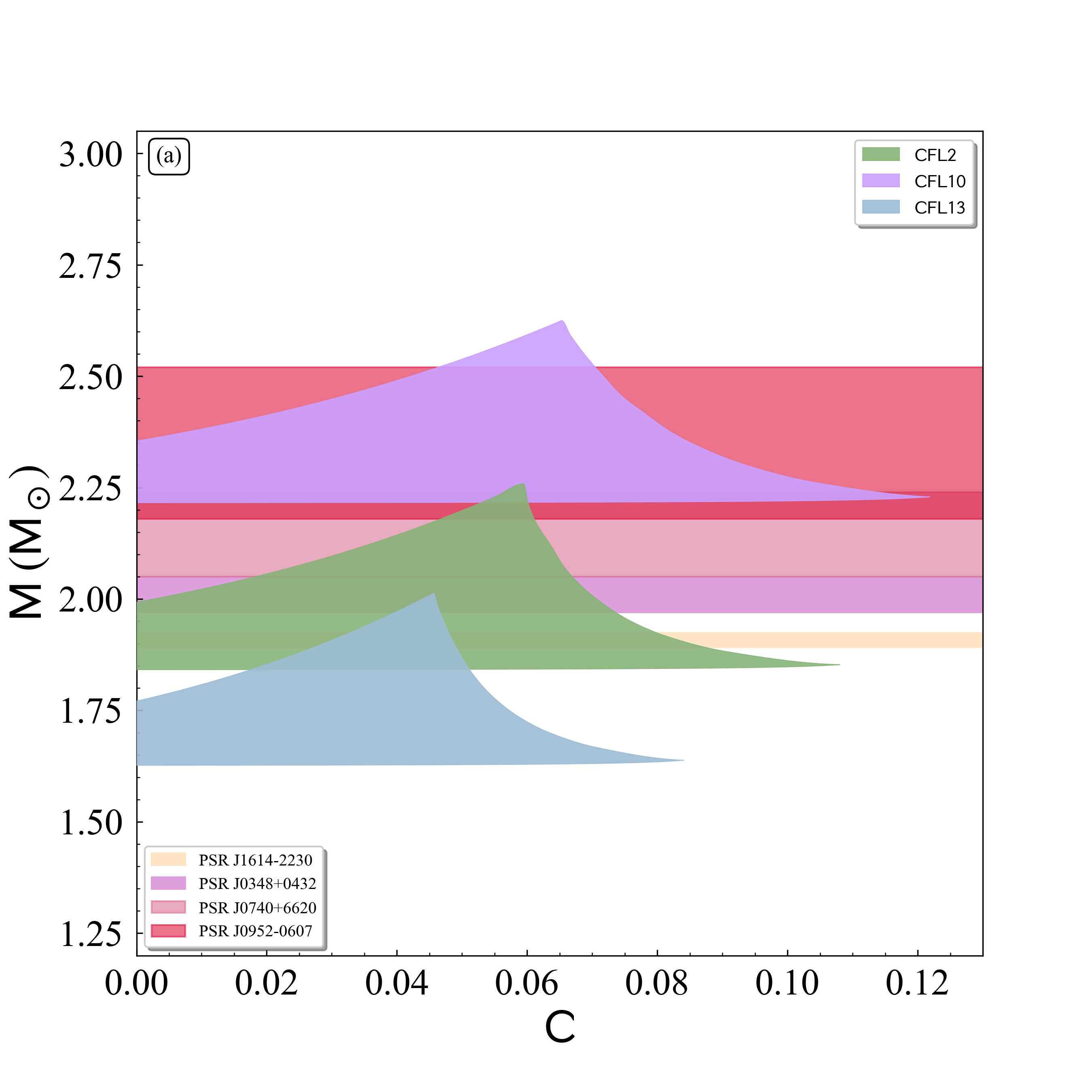}
    ~\includegraphics[width=80mm, height=70mm]{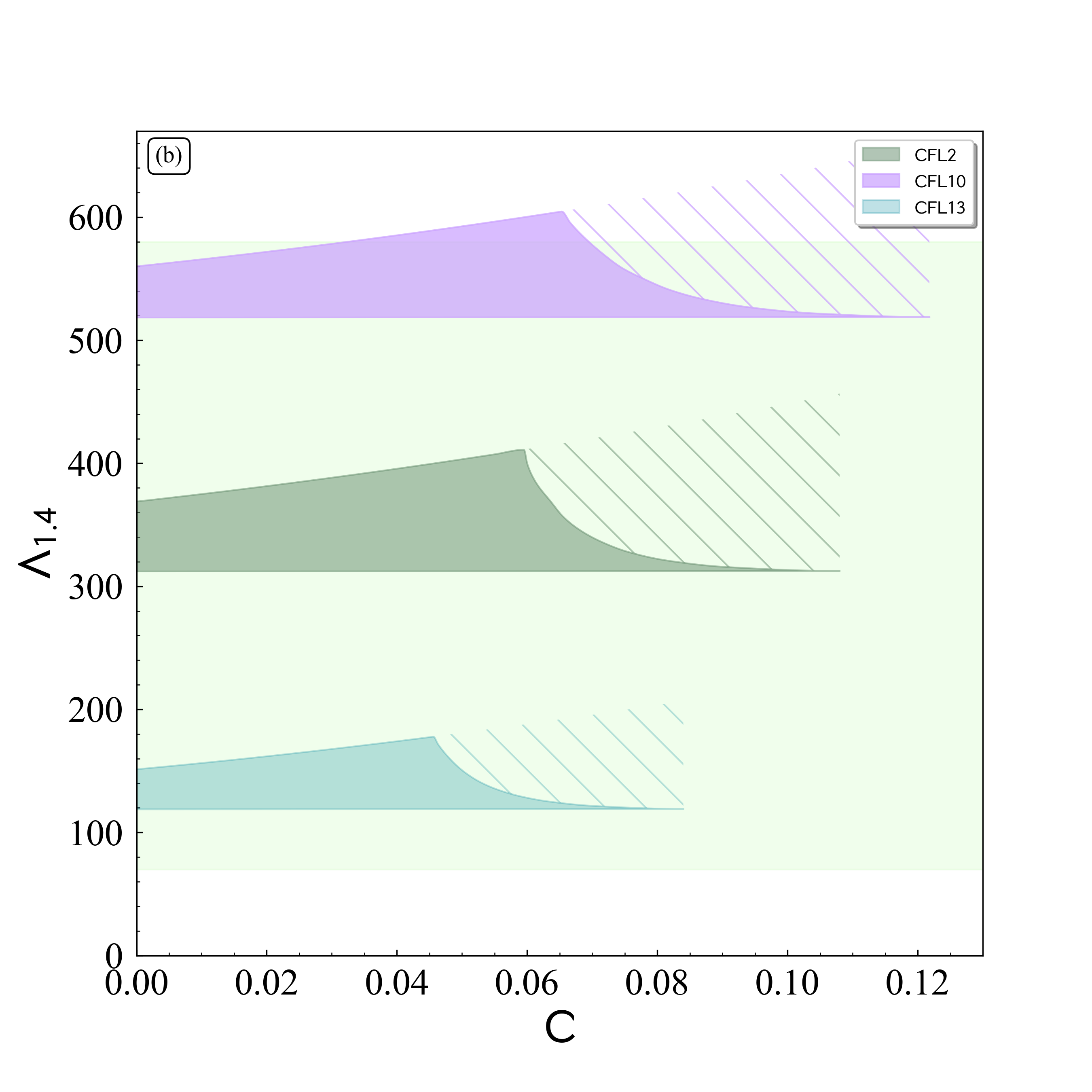}
    \caption{(a) The maximum mass, and (b) the tidal deformability $\Lambda_{1.4}$ corresponding to $\mathrm{1.4\;M_\odot}$ related to the parameter C for the three CFL EoSs. The horizontal shaded areas on the left panel correspond to those of Fig.~\ref{fig:mr_cfl}. The green shaded area on the right panel indicates the constraints from GW170817~\cite{Abbott-2018}, while the colored diagonal lines show the excluded regions from the violation of causality for each EoS.}
    \label{fig:Mmax-L14-C-cfl}
\end{figure*}  

\begin{figure*}[!b]
\centering
    \includegraphics[width=60mm, height=60mm]{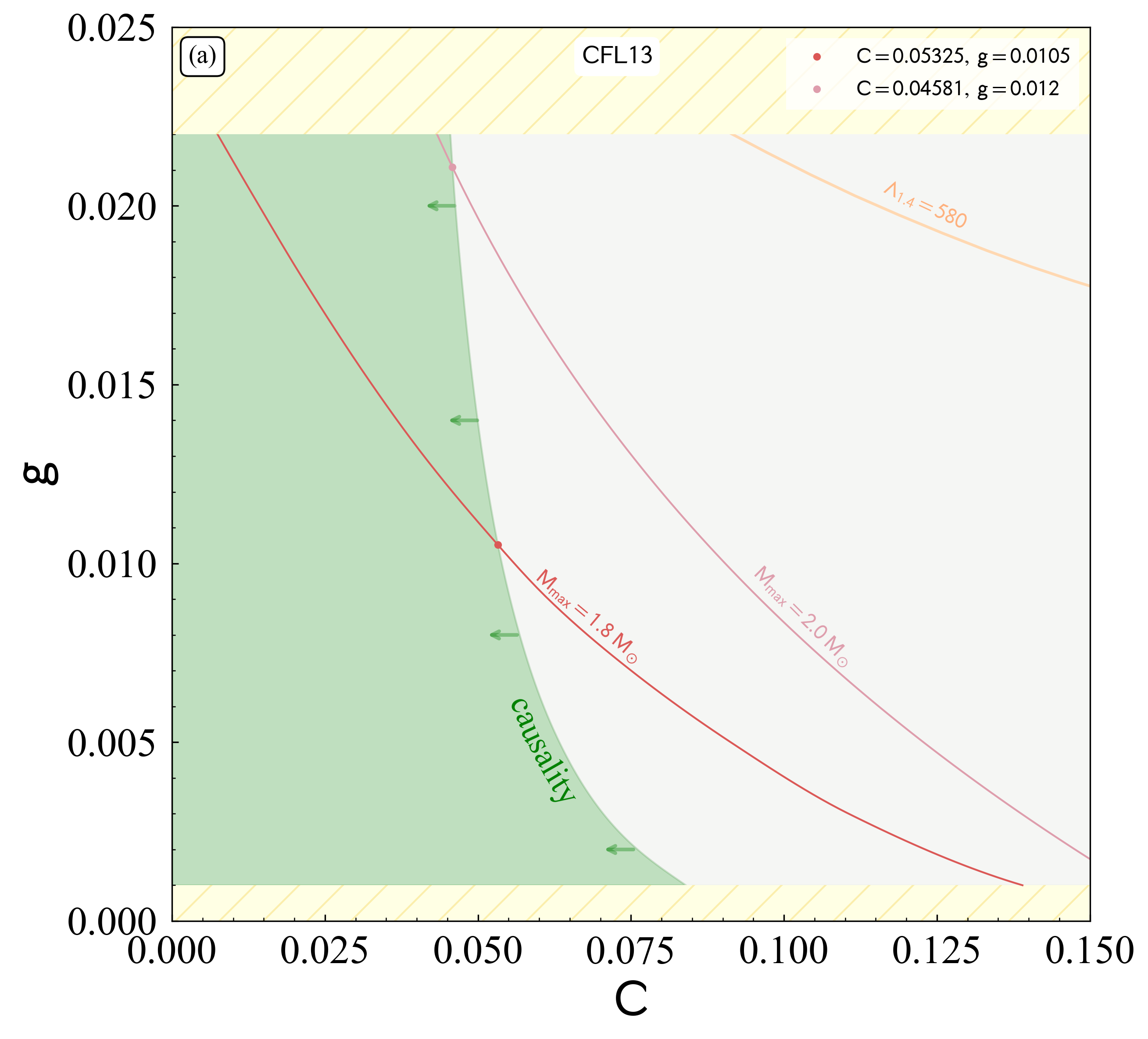}
    ~\includegraphics[width=60mm, height=60mm]{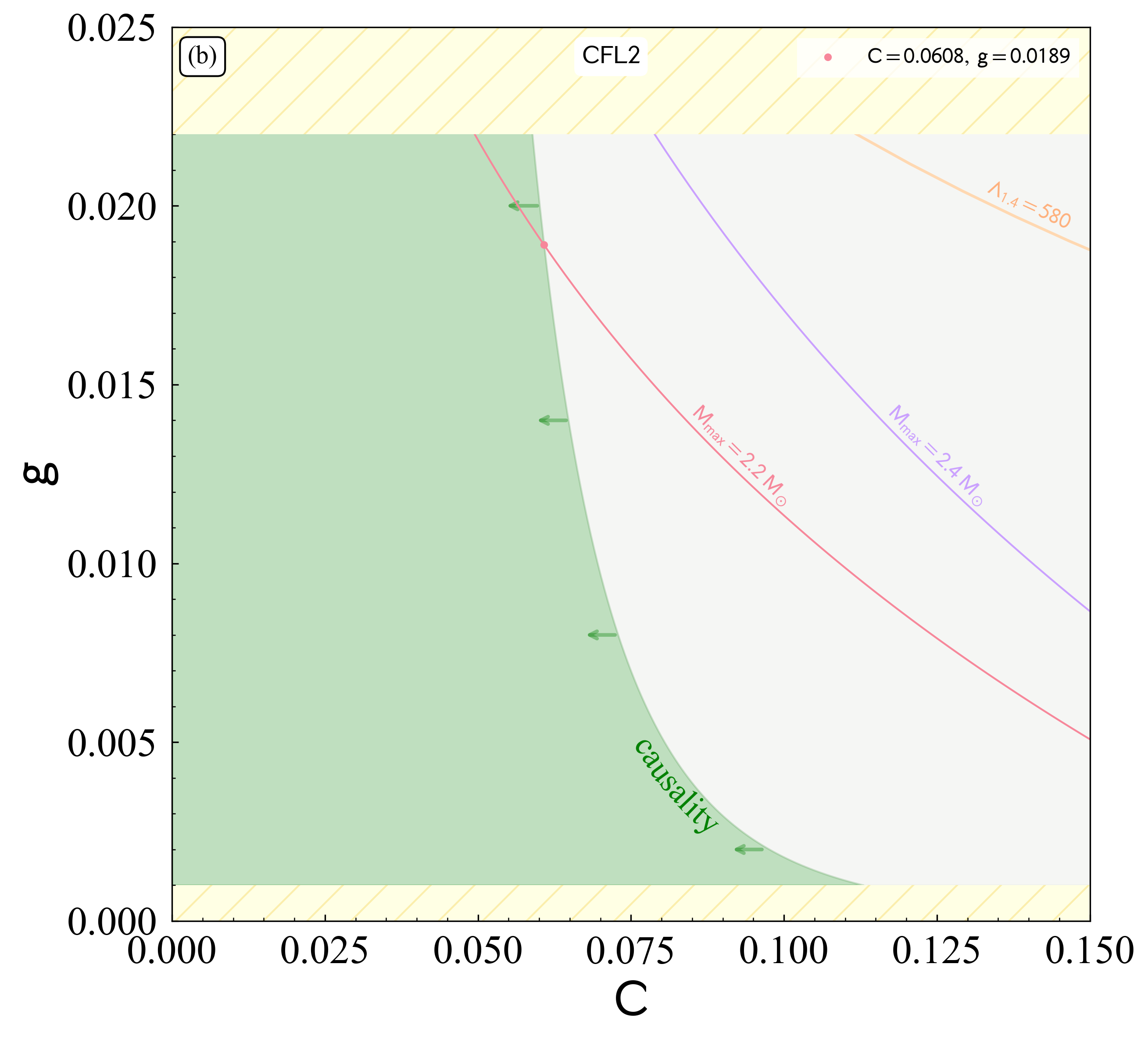}
    ~\includegraphics[width=60mm, height=60mm]{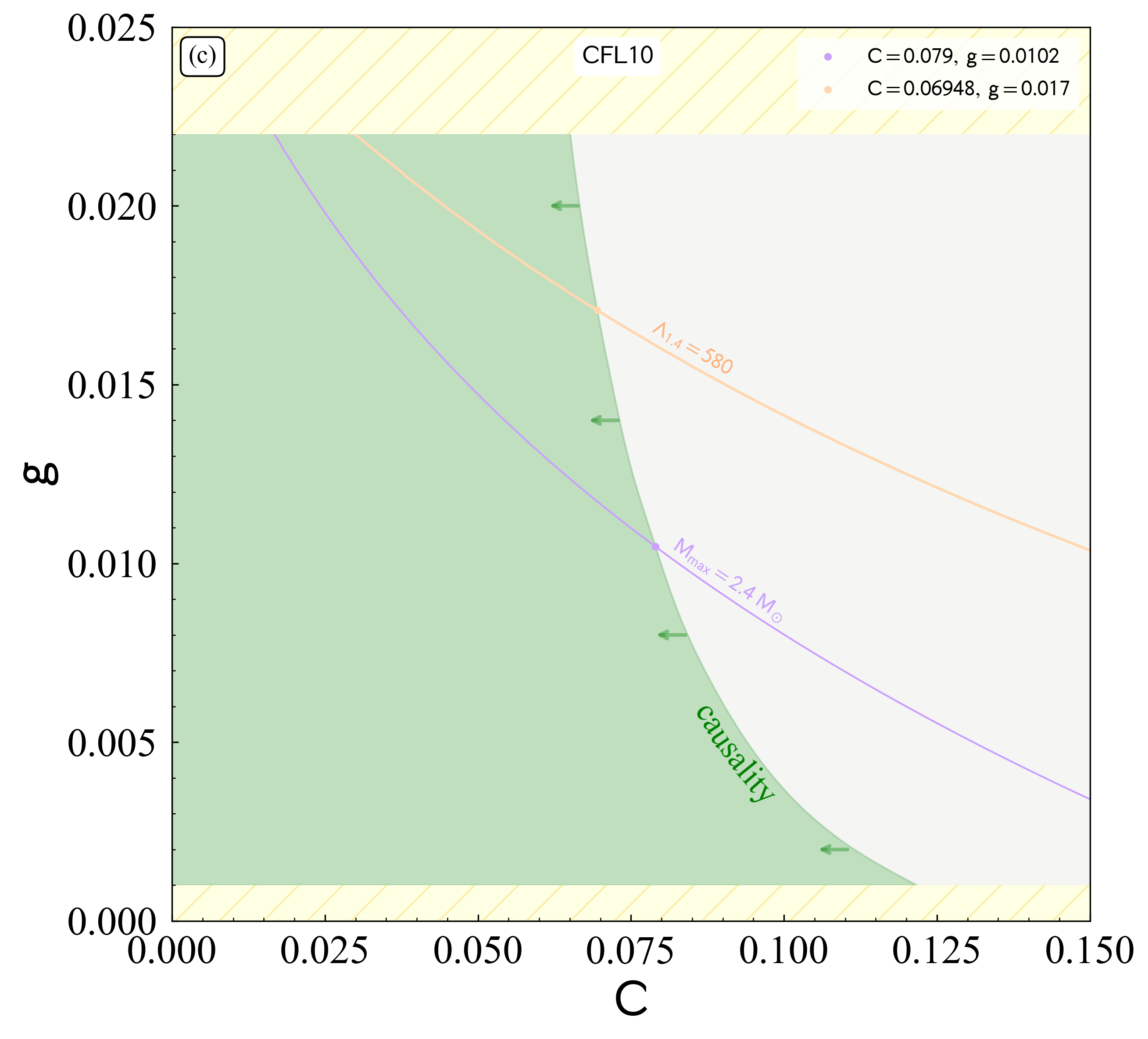}
    \caption{Constraints for $\mathrm{g}$ and $\mathrm{C}$ for (a) the CFL13 quark EoS (soft), the possible upper mass limits  $\mathrm{1.8\;M_{\odot}}$ and $\mathrm{2.0\;M_{\odot}}$ and the dimensionless tidal $\Lambda_{1.4}=580$, (b) the CFL2 quark EoS (intermediate), the possible upper mass limits  $\mathrm{2.2\;M_{\odot}}$ and $\mathrm{2.4\;M_{\odot}}$ and the dimensionless tidal $\Lambda_{1.4}=580$, and (c) the CFL10 quark EoS (stiffer), the possible upper mass limit   $\mathrm{2.4\;M_{\odot}}$ and the dimensionless tidal $\Lambda_{1.4}=580$~\cite{Abbott-2018}. In all panels, the green shaded region indicates the allowed parameter space derived by the non-violation of causality, while the yellow ones indicate the regions where we not include in our study.}   
    \label{fig:g-C-cfl}
\end{figure*} 

In order to take a deeper look on the behavior of the quark EoSs which include the X17 boson, we studied the dependence of their properties related directly to the parameter $\mathrm{C}$, by imposing also available constraints from observations. In Fig.~\ref{fig:Mmax-L14-C-cfl} we display the maximum mass $\mathrm{M_{max}}$ (left panel) and the tidal deformability $\Lambda_{1.4}$ (right panel) related to $\mathrm{C}$ respectively. The blue color corresponds to the CFL13 set of EoSs (soft), the green one to the CFL2 (intermediate), and the purple color corresponds to the CFL10 (stiff). In both panels the colored regions of the EoSs are defined by the non-violation of the causality. In Fig.~\ref{fig:Mmax-L14-C-cfl}(b) the diagonal colored lines indicate the violation of causality. In both $\mathrm{M_{max}-C}$ and $\mathrm{\Lambda_{1.4}-C}$ diagrams, the corresponding peak of each set of EoSs occurs when $\mathrm{g=0.022}$ and $\mathrm{C=C_{max}}$, where $\mathrm{C_{max}}=0.0456$ for CFL10, $\mathrm{C_{max}}=0.0595$ for CFL2, and $\mathrm{C_{max}}=0.0653$ for CFL10. From softer to stiffer EoS this peak shifts to higher values of $\mathrm{C}$, with a corresponding grow on $\mathrm{M_{max}}$ and $\mathrm{\Lambda_{1.4}}$. Across the same value of $\mathrm{g}$, i.e. $\mathrm{g=const.}$ as the $\mathrm{C}$ increases, the effect of the X17 becomes higher; we clarify that the upper limit of each shaded region on the left side of each peak corresponds to $\mathrm{g=0.022}$. The imprint of the high contribution of the X17 to the quark EoSs can be observed clearly from the deviation of the $\mathrm{M_{max}}$ and the $\mathrm{\Lambda_{1.4}}$ from the corresponding value of the initial EoS; e.g. for the CFL2 EoS, $\mathrm{g=0.022}$, and $\mathrm{C=C_{max}}$ it is $\mathrm{\Delta M_{max}\simeq0.417\;M_\odot}$. Regarding the $\mathrm{\Lambda_{1.4}}$, the corresponding deviation for the CFL13 EoS is $\mathrm{\Delta\Lambda_{1.4}\simeq85}$, for the CFL2 EoS it is $\mathrm{\Delta\Lambda_{1.4}\simeq99}$, while for the CFL10 EoS it is $\mathrm{\Delta\Lambda_{1.4}\simeq167}$. Hence, there is an agreement with the gradual change of stiffness. From the Fig.~\ref{fig:Mmax-L14-C-cfl}(b) we notice that all EoSs lie inside the GW170817 accepted region, except some cases with high $\mathrm{g}$ and $\mathrm{C}$ for the CFL10 set of EoSs. So far, the available observational data offered us useful constraints on the two parameters $\mathrm{g}$ and $\mathrm{C}$, but a more detailed view is needed.

For the purpose of the aforementioned examination, we constructed the $\mathrm{g-C}$ parameter space for each set of quark EoS, by applying various upper limits on $\mathrm{M_{max}}$, $\mathrm{\Lambda_{1.4}}$ and including the non-violation of causality. In Fig.~\ref{fig:g-C-cfl} the three parameter spaces are demonstrated; the change on the stiffness (soft to stiff) of the EoS corresponds to the direction from the left to the right panel. This kind of diagrams clarify specifically the role of the two parameters. As we move from softer to stiffer EoS, the range of pair values $\mathrm{(g,C)}$ that satisfy the mentioned constraints increases. Therefore, since the range of values for the coupling constant $\mathrm{g}$ is common for all cases, the stiffer the EoS the higher the value of $\mathrm{C_{max}}$, meaning higher contribution of X17. We underline at this point that across the same set of EoSs the contribution of X17 is characterized firstly by the value of $\mathrm{g}$. The displayed points correspond to the crossed colored curves' values with the causality limit. The curves of $\mathrm{M_{max}(C)}$ and $\mathrm{\Lambda_{1.4}(C)}$ are unique for each set of CFL EoSs. The $\mathrm{M_{max}(C)}$ curves define that the $\mathrm{(g,C)}$ pair of values on the left of each one of these curves do not provided the desired $\mathrm{M_{max}}$. These points are a) $\mathrm{g=0.0105,\;C_{max}=0.05325}$ with $\mathrm{M_{max}=1.8\;M_\odot}$ and $\mathrm{g=0.012,\;C_{max}=0.04581}$ with $\mathrm{M_{max}=2.0\;M_\odot}$ for the CFL13 EoS (soft case), b) $\mathrm{g=0.0189,\;C_{max}=0.0608}$ with $\mathrm{M_{max}=2.2\;M_\odot}$ for the CFL2 EoS (intermediate case), and c) $\mathrm{g=0.0102,\;C_{max}=0.079}$ with $\mathrm{M_{max}=2.4\;M_\odot}$ and $\mathrm{g=0.017,\;C_{max}=0.06948}$ with $\mathrm{\Lambda_{1.4}=580}$ for the CFL10 EoS (stiff case). In the latter case, the observational limit on $\mathrm{\Lambda_{1.4}}$ excludes all the $\mathrm{(g,C)}$ values on the right of its curve. We observe that as we move to stiffer EoS only higher values on $\mathrm{M_{max}}$ can provide further constraints. Concerning the constraints imposed by the $\mathrm{\Lambda_{1.4}}$, these can be useful in stiffer EoSs. 

Concluding, the more soft the equation of state the more limited is the range of parameters, for both hadronic and quark case. In this context, additional observational data concerning the maximum mass as well as more strict upper and even lower limits on $\Lambda_{1.4}$ may lead to much stringent constraints regarding the coupling constant $\mathrm{g}$ and the in-medium effect regulator $\mathrm{C}$.


\section{Concluding Remarks}
We studied the effect of the hypothetical X17 boson on the EoS of neutron star matter as well as on QS and the corresponding bulk properties including mass, radius and tidal deformability.  In particular, we payed attention on two main phenomenological parameters of the X17: a) the coupling constant $\mathrm{g}$ of its interaction with hadrons or quarks, 
and b) the in-medium effects through a regulator $\mathrm{C}$. 
Both are very crucial concerning the contribution on the total energy density and pressure. We suggest that it is possible to provide constraints on these parameters with respect to causality and various possible upper mass limits  and the dimensionless tidal deformability $\Lambda_{1.4}$. Moreover, we found that  the more stiff is the EoS (hadronic or quark), the more indiscernible are the effects on the properties of compact objects. In particular, we found that the effect of the existence of the hypothetical X17 boson, are more pronounced, in the case of QSs, concerning all the bulk properties. This is due mainly on the extra factor 9 both on the energy and pressure contribution to the total ones. It must be emphasized that in the present study,
special attention was paid to maintain the non violation of causality on the EoSs, while systematically taken into account the in-medium effects on the mass of the hypothetical boson (which usually had been omitted
in similar works so far). In addition, an attempt was made to find
possible constraints on the hypothetical X17 boson with the
help of observational data and mainly those derived from
the detection of gravitational waves.  

It would be also of great interest to perform similar calculations in the context of modified gravity theories in order to fully demonstrate the implications of the X17 boson to the properties of various compact objects~\cite{Odintsov:2023ypt,Oikonomou:2023lnh,Oikonomou:2023dgu}.
Finally, it is worth  to notice that the present study can form the framework for similar studies concerning the possible existence of bosons in nuclear matter and their consequences on the structure and basic properties of compact objects. Likely, in this case it will be possible from both terrestrial and astrophysical observations to make the best possible estimate of the properties of these particles, which will concern both individual properties and those related to the interaction with the environment in which they are found. 


\section*{Acknowledgments}
This work is supported by the Czech Science Foundation (GACR Contract  No. 21-24281S)
and by the Hellenic Foundation for Research and
Innovation (HFRI) under the 3rd Call for HFRI PhD Fellowships (Fellowship Number: 5657). One of the authors (Ch.C.M) would like to thank Prof. M.I. Krivoruchenko and  Prof. F. Simkovic for  useful discussion and correspond.

\end{document}